\documentclass[10pt]{article} % For LaTeX2e
\usepackage[accepted]{tmlr}

\usepackage{amsmath,amsfonts,bm}

\def\eqref#1{\textup{(\ref{#1})}}
\def\1{\bm{1}}

\DeclareMathAlphabet{\mathsfit}{\encodingdefault}{\sfdefault}{m}{sl}
\SetMathAlphabet{\mathsfit}{bold}{\encodingdefault}{\sfdefault}{bx}{n}

\usepackage{hyperref}
\hypersetup{hidelinks}
\usepackage{url}
\usepackage{graphicx}
\usepackage{adjustbox}
\usepackage{booktabs}
\usepackage{multirow}
\usepackage[ruled,vlined]{algorithm2e}
\usepackage{xcolor}
\usepackage{tcolorbox}
\usepackage{algorithmic}
\usepackage{colortbl}
\usepackage{comment}
\usepackage{wrapfig}
\usepackage{diagbox}
\usepackage{needspace}
\usepackage{placeins}
\usepackage{float}

\definecolor{myblue}{rgb}{0.85,0.92,0.98}

\newif\ifshowrevisions
\showrevisionsfalse
\newcommand{\revisioncolor}{\ifshowrevisions\color{blue}\fi}
\DeclareRobustCommand{\rev}[1]{{\revisioncolor #1}}
\newenvironment{revision}{\par\begingroup\revisioncolor}{\par\endgroup}
\newif\ifshowaerevisions
\showaerevisionsfalse
\DeclareRobustCommand{\aerev}[1]{{\ifshowaerevisions\color{blue}\fi #1}}
\title{VideoMark: Training-Free Watermarking for 2D-VAE Video Diffusion Models}

\author{%
  \name Hanqian Li\thanks{Equal contribution.} \email hli994@connect.hkust-gz.edu.cn \\
  \addr The Hong Kong University of Science and Technology (Guangzhou) \\
  \name Aiwei Liu\textsuperscript{*,}\thanks{Corresponding authors.} \email liuaiwei20@gmail.com \\
  \addr Tsinghua University \\
  \name Jungang Li\textsuperscript{*} \email lijungang.02@gmail.com \\
  \addr The Hong Kong University of Science and Technology (Guangzhou) \\
  \name Yu Huang \email yhuang3273-c@my.cityu.edu.hk \\
  \addr City University of Hong Kong  \\
  \name Chen Ling \email chen\_ling@zju.edu.cn \\
  \addr Zhejiang University \\
  \name Ye Pan \email ypan477@connect.hkust-gz.edu.cn \\
  \addr The Hong Kong University of Science and Technology (Guangzhou) \\
  \name Shuliang Liu \email shulianglyo@gmail.com \\
  \addr The Hong Kong University of Science and Technology (Guangzhou) \\
  \name Qi Zheng \email qzheng219@connect.hkust-gz.edu.cn \\
  \addr The Hong Kong University of Science and Technology (Guangzhou) \\
  \name Junhao Chen \email junhao@im.sanken.osaka-u.ac.jp \\
  \addr Osaka University \\
  \name Xuming Hu\textsuperscript{$\dagger$} \email xuming@connect.hkust-gz.edu.cn \\
  \addr The Hong Kong University of Science and Technology (Guangzhou)
}

\def\month{11}
\def\year{2026}
\def\openreview{\url{https://openreview.net/forum?id=Ml4KILKIQ9}}

\begin{document}

\maketitle

% Keep the abstract together when the full author list fills the first page.
\noindent\begin{minipage}{\textwidth}
\begin{abstract}
This work presents \textbf{VideoMark}, a training-free watermarking framework for video diffusion models with frame-aligned 2D VAEs. VideoMark embeds pseudorandom error-correcting (PRC) codewords by modulating the signs of Gaussian initialization noise without model retraining. \rev{Under a joint pseudorandomness assumption on the instantiated PRC encoder, the resulting initialization tensor is computationally indistinguishable from the Gaussian reference; exact distributional equality holds for the ideal independent uniform-sign construction.} To support temporally structured video watermarking, we introduce a frame-wise message design together with a Temporal Matching Module (TMM), which recovers frame correspondence under temporal reordering without requiring length-specific watermark layouts. Experiments on ModelScope and I2VGen-XL show that VideoMark achieves accurate watermark extraction on clean generations while maintaining video quality comparable to watermark-free generation. The proposed temporal matching mechanism further improves recovery under frame reordering and enables systematic evaluation under a range of temporal and compression perturbations. We additionally analyze the computational characteristics and practical operating conditions of inversion-based extraction. Overall, VideoMark provides a simple training-free construction for studying pseudorandom watermarking in videos generated by frame-aligned 2D-VAE diffusion models, and establishes a foundation for extending structured noise watermarking to broader video-generation architectures.

\end{abstract}
\end{minipage}\par

\section{Introduction}

Diffusion models have achieved remarkable progress in generating high-quality images and videos \citep{ho2020denoising,ho2022video,sohl2015deep,liu2025javisgpt,rombach2022high,huang2025video}. In parallel, multimodal research is advancing visual reasoning over structured content and complex scenes~\citep{li2026intertab,ling2026seeing,ling2026debias}, as well as continuous video understanding and temporal reasoning~\citep{xun2025rtv,tao2025mosschatv,zhang2026temporal,pan2026egointent}. As synthetic media becomes increasingly realistic, reliable mechanisms for tracing AI-generated content are becoming more important in the context of copyright protection, provenance tracking, and potential misuse \citep{almutairi2022review,zhang2025cohemark,zheng2026visual,pan2025markdiffusion,liu2025vla}. Digital watermarking provides a practical approach for embedding attribution information directly into generated content.

Conventional image and video watermarking methods typically operate as a post-processing step after content generation \citep{luo2023dvmark,zhang2019robust,videoseal}. While effective in many settings, such approaches introduce additional processing and often require dedicated watermark encoders or decoders. An alternative line of work embeds watermarks directly during diffusion generation by exploiting the reversibility of DDIM sampling \citep{song2021denoising}. Methods such as Tree-Ring \citep{wen2023tree}, Gaussian Shading \citep{yang2024gaussian}, and PRC-Watermark \citep{gunn2025an} encode information through structured modifications of the initial Gaussian noise, enabling training-free watermark injection without an additional embedding network.

Extending initialization-noise watermarking from images to videos introduces an additional temporal dimension. In a video, watermark messages must not only be recoverable from individual frames, but also remain identifiable when temporal order changes. Moreover, a practical decoding mechanism should accommodate different video lengths without requiring a separately predefined watermark layout for each sequence length. These considerations make temporal indexing and synchronization important components of video watermark design.

In this work, we study this problem in frame-aligned video diffusion models built upon 2D VAEs, where individual video frames maintain an explicit correspondence with latent representations. This setting enables a direct extension of frame-wise initialization-noise watermarking while exposing a key challenge absent in single-image generation: independently embedded frame messages must still preserve sufficient temporal structure for reliable sequence-level recovery.

Based on this observation, we formulate two design objectives. First, watermark embedding should preserve the statistical properties of the diffusion initialization noise so that generation quality is minimally affected. Second, frame-level watermark messages should carry recoverable temporal information, allowing their correspondence to be reconstructed even when the observed frame order changes.

To address these objectives, we introduce \textbf{VideoMark}, a training-free watermarking framework for frame-aligned 2D-VAE video diffusion models. VideoMark builds on PRC-Watermark~\citep{gunn2025an} and pseudorandom error-correcting (PRC) codes~\citep{christ2024pseudorandom} to modulate initialization-noise signs. \rev{Its distributional guarantee is conditional computational indistinguishability of the joint latent tensor, assuming joint pseudorandomness of repeated PRC encodings under the same secret key (Appendix~\ref{sec:joint_prc_guarantee}).} Rather than assigning a fixed watermark layout to a particular video length, VideoMark constructs a shared watermark message sequence and samples a random starting position for each generated video. The first frame is associated with the sampled position, while subsequent frames follow sequentially in the same message space. This design provides frame-specific watermark messages while retaining relative temporal structure across the video.

For watermark extraction, we introduce a Temporal Matching Module (TMM) that localizes decoded frame messages within the shared message sequence and reconstructs their relative ordering. As a result, temporal correspondence can be recovered without registering a separate decoding layout for each possible video length. The resulting framework combines frame-wise PRC initialization, random-start message registration, and temporal index recovery into a unified video watermarking pipeline. \rev{The video-specific contribution is this combination of message assignment, registration, and temporal recovery; the nearest-neighbor matching primitive and frame-level PRC embedding are not new coding algorithms.}

We evaluate VideoMark on ModelScope and I2VGen-XL. Experiments show that the method achieves high decoding accuracy on unmodified generations while maintaining video quality comparable to watermark-free generation according to VBench \citep{huang2024vbench}. We further study its behavior under video-length variation, temporal transformations, compression, extraction mismatch, latent-space perturbations, and different extraction settings. These experiments show that the temporal matching mechanism is particularly effective for frame reordering, while more destructive temporal or latent transformations can degrade watermark recovery. We additionally analyze the computational cost and operating conditions of inversion-based extraction to provide a clear characterization of the settings in which the method is effective.

Overall, VideoMark is intended as a scoped study of frame-wise pseudorandom watermarking for short videos generated by frame-aligned 2D-VAE diffusion models. Rather than assuming general robustness to arbitrary video post-processing, we focus on understanding how structured initialization-noise watermarks can be extended to the temporal domain and how temporal indexing can improve recovery under frame-level reordering.

In summary, our contributions are as follows:
\begin{itemize}
\item We introduce a training-free frame-wise PRC watermarking construction for frame-aligned 2D-VAE video diffusion models that embeds information directly into the initialization noise without training an auxiliary embedding network.

\item We propose a random-start message assignment strategy and a Temporal Matching Module that recover relative temporal correspondence without requiring a length-specific watermark layout.

\item We provide a systematic empirical analysis across video length, temporal transformations, compression, latent-space perturbations, extraction mismatch, and computational cost, characterizing both the effective operating regime and the practical limits of inversion-based video watermark recovery.
\end{itemize}

\section{Related Work}

\subsection{Video Diffusion Models}

Diffusion models learn to reverse a progressive noising process \citep{sohl2015deep,ho2020denoising}. Video diffusion models extend this framework with temporal modeling to generate coherent frame sequences \citep{ho2022video}, while latent diffusion improves generation efficiency by operating in a compressed space \citep{rombach2022high,blattmann2023stable}. DDIM sampling \citep{song2021denoising} further enables efficient sampling and approximate inversion, which is useful for initialization-noise watermarking.

Video generation mainly includes Text-to-Video \citep{wang2023modelscope,huang2024vbench} and Image-to-Video \citep{2023i2vgenxl,blattmann2023stable}. In this work, we focus on frame-aligned 2D-VAE video diffusion models, where individual frames maintain explicit correspondence with latent representations.

\subsection{Video Watermarking}

Video watermarking methods can broadly be divided into post-processing and generation-integrated approaches. Post-processing methods embed watermarks after generation, often using learned embedding and extraction networks \citep{videoseal,luo2023dvmark,zhang2019robust,revmark}.

Generation-integrated methods instead embed watermark information during synthesis. Video Signature~\citep{huang2025video} partially fine-tunes the latent decoder, using perturbation-aware suppression and temporal alignment to embed watermarks while preserving spatial quality and temporal coherence. VideoShield \citep{hu2025videoshield} extends Gaussian Shading \citep{yang2024gaussian} to video diffusion through initialization-noise modulation. Other recent approaches include Safe-Sora \citep{su2025safesora}, SPDMark \citep{fares2025spdmark}, and SIGMark \citep{zhu2026sigmark}, which explore different watermark representations, model-access assumptions, and video architectures.
VideoMark focuses on a complementary setting: training-free watermarking through pseudorandom initialization-noise modulation for frame-aligned 2D-VAE video diffusion models, together with frame-wise message assignment and temporal matching.

Watermarking also extends to vision-language generation. VISA-Mark~\citep{zheng2026visual} uses a lightweight prefix-tuner to estimate visual evidence and guide token-level watermarking in LVLM outputs. This addresses visual grounding in generated text, whereas VideoMark embeds recoverable messages in generated video frames. Complementary to these algorithmic developments, MarkDiffusion~\citep{pan2025markdiffusion} provides a unified toolkit for implementing, visualizing, and evaluating latent-diffusion watermarks.

\section{Preliminaries}
\subsection{Diffusion Models}
Diffusion models generate content through an iterative denoising process. Given a noise schedule $\{\beta_t\}_{t=1}^T$, the forward process gradually adds noise to data $\mathbf{x}_0$:
\begin{equation}
q(\mathbf{x}_t|\mathbf{x}_{t-1}) = \mathcal{N}(\mathbf{x}_t; \sqrt{1-\beta_t}\mathbf{x}_{t-1}, \beta_t\mathbf{I})
\end{equation}
The reverse process is learned to gradually denoise from $\mathbf{x}_T \sim \mathcal{N}(0, \mathbf{I})$ to generate content. While DDPM \citep{ho2020denoising} introduces stochasticity in each denoising step, DDIM \citep{song2021denoising} provides an approximately invertible deterministic sampling process:
\begin{equation}
\begin{split}
\mathbf{x}_{t-1} =\ & \sqrt{\alpha_{t-1}} \left( 
    \frac{\mathbf{x}_t - \sqrt{1-\alpha_t}\,\boldsymbol{\epsilon}_\theta(\mathbf{x}_t, t)}{\sqrt{\alpha_t}} 
\right) \\
& + \sqrt{1-\alpha_{t-1}}\,\boldsymbol{\epsilon}_\theta(\mathbf{x}_t, t)
\end{split}
\end{equation}

This deterministic reversibility enables control over the generation process through manipulation of the initial noise.
\subsection{Pseudorandom Codes (PRC)}
A PRC maps messages to computationally pseudorandom codewords. We use the PRC-Watermark implementation~\citep{gunn2025an}, based on the PRC framework of \citet{christ2024pseudorandom}. The underlying theory relies on computational hardness assumptions, including Learning Parity with Noise (LPN). \rev{Our multi-frame guarantee additionally requires joint pseudorandomness for the instantiated encoder, parameters, and message process under repeated use of a secret key; it is not an unconditional claim for every practical parameter choice.}

The PRC framework consists of three core algorithms:
\begin{itemize}
    \item $\text{KeyGen}(n, m, \text{fpr}, t) \rightarrow \text{key}$: Generates a key for encoding $m$-bit messages into $n$-bit codewords with sparsity parameter $t$
    \item $\text{Encode}(\text{key}, \mathbf{m}) \rightarrow \mathbf{c}$: Maps message $\mathbf{m}$ to codeword $\mathbf{c} \in \{-1,1\}^n$
    \item $\text{Decode}(\text{key}, \mathbf{s}) \rightarrow \mathbf{m}$ or $\emptyset$: Recovers message from potentially corrupted signal $\mathbf{s} \in [-1,1]^n$
\end{itemize}

Our implementation retains soft confidence values during PRC decoding (see supplementary materials for details).

\subsection{Generative Video Watermarking and Threat Model}
Diffusion-based video watermarking involves three key functions in the watermarking process:

1. \textbf{Generation}: $V = \mathcal{G}(m, k)$, where $\mathcal{G}$ generates a watermarked video $V$ by embedding message $m$ using secret key $k$ during the diffusion process.

2. \textbf{Decoding}: $\hat{m} = \mathcal{D}_{PRC}(V)$, where $\mathcal{D}_{PRC}$ extracts the watermark message $\hat{m}$  from the given video $V$.

3. \textbf{Detection}: $\{p, d\} = \text{Detect}(m, \hat{m})$, where $\text{Detect}$ compares the original message $m$ with the decoded message $\hat{m}$. This function outputs a p-value $p$ and a boolean decision $d$ indicating whether the distance between $m$ and $\hat{m}$ is significantly smaller than that between $m$ and a random message.

We consider active adversaries who may deliberately apply transformations to remove, distort, or invalidate the watermark. Our current implementation targets diffusion models with a 2D VAE, where one latent frame corresponds to one output frame. Extraction is non-blind with respect to the model and watermark resources: it requires the corresponding model and VAE, scheduler, inversion-step and guidance configurations, preprocessing parameters, private PRC key, and registered master message sequence. \rev{It does not require the original generation prompt or original conditioning image. Throughout the reported experiments, T2V inversion uses an empty prompt; I2V inversion uses the fixed literal string \texttt{\char34 none\char34} and the first frame of the video being verified as image conditioning.} We evaluate both matched extraction and moderate configuration mismatch. Temporally compressed causal-3D-VAE models require a redesigned grouping and alignment mechanism and are outside the scope of this implementation.
\begin{itemize}
    \item \textbf{Temporal Attacks}: Frame dropping, insertion, or reordering, which disrupts the temporal continuity and alters the original frame sequence, leading to misalignment in the temporal structure and potentially degrading the integrity and decodability of the embedded watermark.
    
    \item \textbf{Spatial Attacks}: Frame-wise manipulations such as Gaussian blurring, colour jittering, and resolution compression, which aim to distort the watermark signal by degrading the visual content of individual frames.
\end{itemize}
Our design specifically targets frame-wise corruption and temporal reordering. Operations that irreversibly discard or resample watermark-bearing frames, including aggressive trimming, frame-rate conversion, or latent smoothing, can substantially reduce recovery accuracy; Section~\ref{sec:expanded_robustness} characterizes these boundaries.

\section{Proposed Method}
\label{Proposed_Method}
\begin{figure*}[t]
    \centering
    \includegraphics[width=\textwidth]{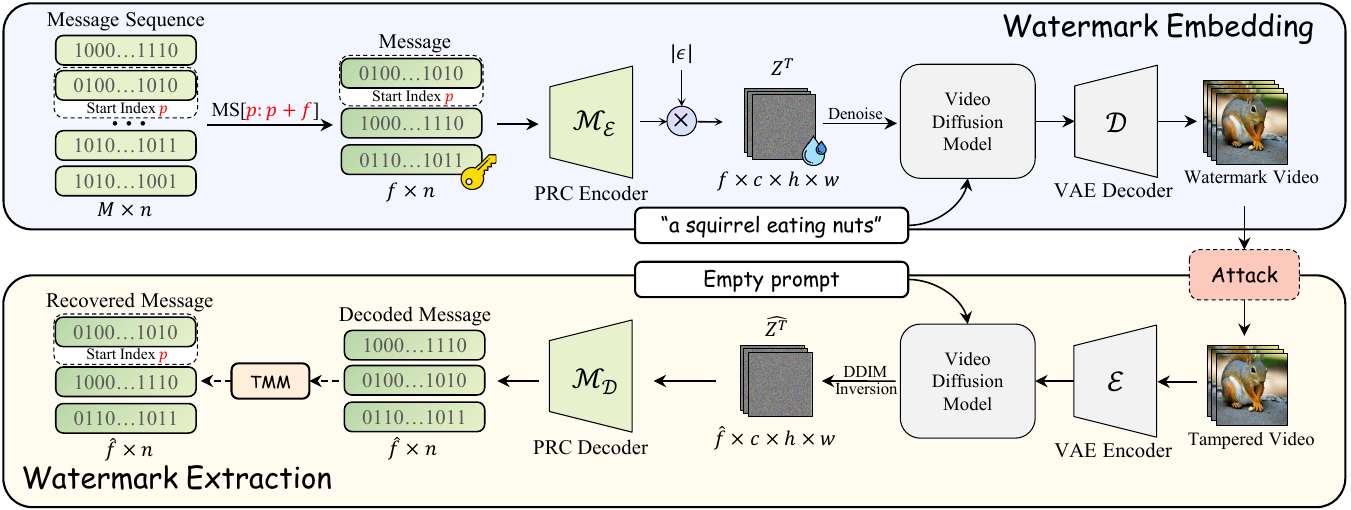}
    \caption{
    The overall framework of VideoMark. During the watermark embedding phase, $\epsilon$ denotes the standard Gaussian noise sampled randomly. \rev{The empty-prompt inversion shown here is the T2V configuration. I2V inversion uses the fixed literal string \texttt{\char34 none\char34} and the first frame of the video being verified, without the original generation prompt or conditioning image.}
}

    \label{fig:pipeline}
\end{figure*}

In this section, we describe the proposed training-free watermarking method for video diffusion models. Section~\ref{subsec:generation} details watermark generation, and Section~\ref{subsec:extraction} presents extraction.

\subsection{Watermark Generation}
\label{subsec:generation}

\begin{algorithm}[t]
\caption{Watermarked Video Generation}
\label{alg:generation}
\SetKwInOut{Input}{Input}
\SetKwInOut{Output}{Output}

\Input{PRC key $k$, registered master sequence $M=[m^{(1)},\ldots,m^{(L)}]$, number of frames $f$, channels $c$, height $h$, width $w$, diffusion model $\mathcal{M}$, VAE decoder $\mathcal{D}$}
\Output{Watermarked video $V$}
\BlankLine

Sample $p \sim \operatorname{Uniform}\{1,\ldots,L-f+1\}$\;

\For{$i = 1$ \KwTo $f$}{
    Assign frame message $m_i \gets m^{(p+i-1)}$\;
    
    Encode $m_i$ \rev{with key $k$ and fresh randomness} into PRC codeword $c_i \in \{-1, 1\}^{c \times h \times w}$\;
    
    Sample $\epsilon_i \sim \mathcal{N}(0, 1) \in \mathbb{R}^{c \times h \times w}$\;
    
    Compute $\hat{\epsilon}_i \gets c_i \cdot |\epsilon_i|$\;
}

Denoise $\hat{\epsilon}$ using diffusion model $\mathcal{M}$ and decode with VAE decoder $\mathcal{D}$ to obtain video $V$\;

\Return $V$\;
\end{algorithm}

VideoMark initializes each frame with PRC-modulated pseudorandom Gaussian noise, followed by DDIM denoising \citep{song2021denoising} and VAE decoding \citep{vae}. An extended message list with a random start index provides per-video diversity and avoids a separate registration layout for every supported frame count.

VideoMark generates frame-specific pseudorandom initializations using randomized PRC encoding. For a video with $f$ frames, dimensions $c \times h \times w$ (channels, height, width), and a $b$-bit message $m_i'$ per frame, the process is as follows. For each frame $i \in \{1, \ldots, f\}$, we sample Gaussian noise $\boldsymbol{\epsilon}_i \sim \mathcal{N}(0, \mathbf{I}) \in \mathbb{R}^{c \times h \times w}$. Using a PRC key $k$, we encode $m_i'$ to obtain a codeword $\mathbf{c}_i = \text{Encode}(k, m_i') \in \{-1, 1\}^{c \times h \times w}$. \rev{The key is shared across frames, while each encoding invocation uses fresh randomness independent of the sampled Gaussian magnitudes.} The watermarked noise is formally defined as:
\[
\hat{\boldsymbol{\epsilon}}_i = \mathbf{c}_i \cdot |\boldsymbol{\epsilon}_i|,
\]
where $\mathbf{c}_i$ modulates the sign of $\boldsymbol{\epsilon}_i$, preserving its magnitude. Appendix~\ref{sec:joint_prc_guarantee} distinguishes exact Gaussian equality for independent uniform signs from conditional computational indistinguishability for the implemented PRC signs, including dependencies across frames. The noise sequence $\hat{\boldsymbol{\epsilon}} = [\hat{\boldsymbol{\epsilon}}_1, \ldots, \hat{\boldsymbol{\epsilon}}_f]$ is denoised using a DDIM diffusion model $\mathcal{M}$. For each frame, DDIM iterates over $T$ steps:
\begin{equation}
\begin{split}
\hat{\boldsymbol{z}}_i^{(t-1)} =\ & \sqrt{\alpha_{t-1}} \left(
\frac{\hat{\boldsymbol{z}}_i^{(t)} - \sqrt{1-\alpha_t} \,\boldsymbol{\epsilon}_\theta(\hat{\boldsymbol{z}}_i^{(t)}, t)}{\sqrt{\alpha_t}} 
\right) \\
& + \sqrt{1-\alpha_{t-1}}\, \boldsymbol{\epsilon}_\theta(\hat{\boldsymbol{z}}_i^{(t)}, t),
\end{split}
\end{equation}

with $\hat{\boldsymbol{z}}_i^{(T)} = \hat{\boldsymbol{\epsilon}}_i$, producing latent $\hat{\boldsymbol{z}}_i^{(0)}$. The VAE decoder $\mathcal{D}$ generates the watermarked video $V = [\mathcal{D}(\hat{\boldsymbol{z}}_1^{(0)}), \ldots, \mathcal{D}(\hat{\boldsymbol{z}}_f^{(0)})]$. 

To support videos with different frame counts without requiring length-specific watermark layouts, we construct an extended message sequence $M=[m^{(1)},\ldots,m^{(L)}]$, where $L>f_{\max}$. For a video with $f$ frames, we sample a random starting position $p\sim\operatorname{Uniform}\{1,\ldots,L-f+1\}$ and assign consecutive messages as $m_i'=m^{(p+i-1)}$. This random-start strategy provides each frame with a distinct watermark message while preserving their relative temporal order in the shared sequence, enabling temporal localization during extraction. The overall framework and detailed implementation are illustrated in Fig.~\ref{fig:pipeline} and Algorithm~\ref{alg:generation}, respectively.

\subsection{Watermark Extraction}
\label{subsec:extraction}

Our watermark extraction process consists of decoding, detection, and temporal recovery. Appendix~\ref{sec:extraction_details} discusses their distinct outcomes and limitations.

\subsubsection{Decoding Function} $\hat{m} = \mathcal{D}_{PRC}(V)$ extracts the embedded message from a watermarked video $V$ with $f$ frames. We first recover the approximate initial noise for each frame using the DDIM inverse process:
\begin{equation}
\tilde{\boldsymbol{\epsilon}}_i = \mathcal{M}^{-1}(V_i), \quad  i \in \{1,\ldots,f\}.
\end{equation}
We retain the continuous confidence of every recovered latent coefficient rather than applying a hard sign threshold. With scale parameter $\sigma$, the soft observation and decoded frame message are
\begin{equation}
\mathbf{s}_i = \operatorname{erf}\!\left(\tilde{\boldsymbol{\epsilon}}_i/\sigma\right),
\qquad
\hat{m}_i = \operatorname{PRC.Decode}_{\mathrm{BP}}(k,\mathbf{s}_i),
\end{equation}
where $\mathbf{s}_i\in[-1,1]^{c\times h\times w}$ is passed directly to the BP-OSD soft decoder. The decoded sequence is $\hat{m}=[\hat{m}_1,\ldots,\hat{m}_f]$; Appendix~\ref{prc_watermark} provides the decoder details.

\subsubsection{Detection Function} $\{p,d\}=\operatorname{Detect}(m,\hat m)$ compares two equal-length, $b$-bit frame messages using normalized Hamming distance,
\begin{equation}
 d_H(m,\hat m)=\frac{1}{b}\sum_{q=1}^{b}\mathbb{I}[m_q\neq\hat m_q].
\end{equation}
To assess significance, we draw $N$ random $b$-bit messages $r^1,\ldots,r^N$ and compute
\begin{equation}
\label{eq:mc_detection}
 p=\frac{1+\sum_{n=1}^{N}\mathbb{I}[d_H(r^n,\hat m)\leq d_H(m,\hat m)]}{N+1},
 \qquad d=\mathbb{I}[p<\tau].
\end{equation}
We use $\tau=0.005$ by default; Appendix~\ref{sec:extraction_details} discusses Monte Carlo resolution and Appendix~\ref{sec:detection_reliability} reports detection controls and threshold sensitivity.
\aerev{For video-level detection, each Monte Carlo reference trial repeats the same TMM search and video-level aggregation as the observed messages, so the reference distribution includes the effect of matching (Appendix~\ref{sec:tmm_calibration}).}

\textbf{Temporal Matching Function.} Let $M=[m^{(1)},\ldots,m^{(L)}]$ be the registered master sequence. For every decoded frame message $\hat m_j$, TMM independently recovers its most likely master-sequence index:
\begin{equation}
 I_j=\arg\min_{\ell\in\{1,\ldots,L\}}d_H\!\left(m^{(\ell)},\hat m_j\right).
\end{equation}
The recovered pairs are ordered by their indices,
\begin{equation}
 \mathcal{T}(\hat m)=\big[(I_j,\hat m_j)\big]_{j\in\operatorname{argsort}(I_1,\ldots,I_f)}.
\end{equation}
TMM restores the order of decoded frame messages. Appendix~\ref{sec:extraction_details} gives an example and discusses temporal representations, while Appendix~\ref{sec:component_ablation} evaluates its contribution.

\section{Experiments}
\label{experiments}
\subsection{Experimental Setting}
\subsubsection{Implementation Details} We evaluate text-to-video (T2V) and image-to-video (I2V) generation with the 2D-VAE-based ModelScope (MS) \citep{wang2023modelscope} and I2VGen-XL \citep{2023i2vgenxl}. Primary experiments use 16 frames at $512\times512$ resolution, with 25 and 50 default inference/inversion steps for T2V and I2V, respectively. Unless stated otherwise, clean extraction decodes 512 verification-code bits per watermarked frame. Spatial robustness comparisons use 32 bits per frame over 16 frames (512 total bits), matching VideoShield's total evaluation payload; we separately report robustness at 512 bits per frame and payload--robustness curves in Section~\ref{sec:payload_robustness}. \rev{T2V inversion uses an empty prompt; I2V uses the literal string \texttt{\char34 none\char34} and the first observed video frame. These conditions apply throughout, including the extraction-mismatch experiments.} The default detection threshold is $\tau=0.005$. Original p-values use $N=1000$ Monte Carlo references; \rev{supplementary video-detection results use $N=20{,}000$, with threshold sensitivity reported separately.} We run all experiments on an NVIDIA Tesla A800 80GB GPU.

\subsubsection{Baselines} We compare with RivaGAN~\citep{zhang2019robust}, REVMark~\citep{revmark}, VideoSeal~\citep{videoseal}, and VideoShield~\citep{hu2025videoshield}. Following their supported configurations, we use payloads of 32, 96, 96, and 512 bits, respectively.
VideoShield is the directly comparable in-generation baseline in our experiments; the other baselines use trained post-processing encoders and decoders.

\subsubsection{Datasets} We select 50 prompts from the VBench test set~\citep{huang2024vbench}, with 10 prompts each for animals, humans, plants, scenery, and vehicles. For T2V, we generate four videos per prompt. For I2V, Stable Diffusion 2.1~\citep{rombach2022high} first generates a conditioning image for each prompt. The primary evaluation contains 200 videos per task. A separate detectability experiment uses 8,000 watermarked and 8,000 watermark-free videos.

\subsubsection{Metrics} We assess detectability, robustness, and output quality, the three evaluation dimensions also organized by MarkDiffusion~\citep{pan2025markdiffusion}. We report bit accuracy and the VBench Subject Consistency, Background Consistency, Motion Smoothness, and Image Quality metrics. Complementary GPT-4o evaluation receives complete ordered frame sequences. Appendix~\ref{MLLM_for_video_quality} reports its protocol, per-sample winner counts, mean scores, and limitations.

\subsection{Main Results} 
Table~\ref{tab:main_result} summarizes extraction accuracy, video quality, and robustness to temporal and spatial manipulations.
\subsubsection{Extraction} The ``Extraction'' columns report watermark capacity and bit accuracy.
For I2V, the input conditioning frame is not generated by the diffusion process and is therefore excluded; all reported I2V results apply to the 15 generated frames. VideoMark achieves bit accuracies of 1.000 and 0.997 when decoding raw payloads of $512\times16$ and $512\times15$ bits, respectively. These values describe the frame-wise watermark channel; the number of identities supported by a deployment also depends on its message-assignment and registration protocol. They must not be conflated with the matched 512-total-bit setting used in the spatial robustness columns.
Under clean extraction, VideoMark is comparable to VideoShield and exceeds VideoSeal and REVMark.
%\begin{comment}
\begin{table*}[t]
\centering
\caption{Main results of \textit{VideoMark}. ``Decoded Bits'' is the raw payload recovered under clean extraction; deployment-level identity capacity depends on the message-assignment and registration protocol. Spatial robustness uses a matched total payload of 512 bits (32 bits/frame for VideoMark); all columns report bit accuracy except Video Quality. Robustness at 512 bits per frame is reported in Appendix~\ref{sec:payload_robustness}.}
\begin{adjustbox}{width=\textwidth, center}
\small
\begin{tabular}{c l cc ccccc cccc cccc}
\toprule
\multirow{2}{*}{Model} & \multirow{2}{*}{Method} & \multicolumn{2}{c}{\textbf{Extraction}} &
\multicolumn{5}{c}{\textbf{Video Quality}} &
\multicolumn{4}{c}{\textbf{Temporal Tampering}} &
\multicolumn{4}{c}{\textbf{Spatial Tampering}} \\
\cmidrule(lr){3-4} \cmidrule(lr){5-9} \cmidrule(lr){10-13} \cmidrule(lr){14-17}
 & & Decoded Bits & Acc. & SC & BC & MS & IQ & Avg. & Swap & Insert & Drop & Avg. & G. Blur & C. Jitter & R. Comp. & Avg. \\
\midrule
\multirow{5}{*}{MS} 
    & RivaGAN & 32 & 0.994 & 0.922 & 0.951 & 0.960 & 0.648 & 0.870 & 0.930 & 0.919 & 0.930 & 0.926 & 0.919 & 0.939 & 0.783 & 0.880 \\
    & REVMark & 96 & 0.996 & 0.943 & 0.960 & 0.972 & 0.450 & 0.831 & 0.992 & - & - & - & 0.987 & 0.765 & 0.508 & 0.753 \\
    & VideoSeal & 96 & 0.964 & \underline{0.950} & 0.959 & \textbf{0.977} & 0.679 & 0.891 & 0.960 & \underline{0.960} & \underline{0.961} & 0.961 & 0.964 & 0.964 & 0.565 & 0.831 \\
    & VideoShield & 512 & \textbf{1.000} & 0.949 & \textbf{0.962} & \textbf{0.977} & \underline{0.689} & \underline{0.894} & \textbf{1.000} & - & - & - & \textbf{1.000} & \textbf{1.000} & \underline{0.999} & \underline{1.000} \\
    & \cellcolor{myblue}\textit{VideoMark} & \cellcolor{myblue}{512$\times$16} & \cellcolor{myblue}\textbf{1.000} &
      \cellcolor{myblue}\textbf{0.951} & \cellcolor{myblue}\underline{0.961} & \cellcolor{myblue}\textbf{0.977} & \cellcolor{myblue}\textbf{0.692} & \cellcolor{myblue}\textbf{0.895} &
      \cellcolor{myblue}\textbf{1.000} & \cellcolor{myblue}\textbf{1.000} & \cellcolor{myblue}\textbf{1.000} & \cellcolor{myblue}\textbf{1.000} &
      \cellcolor{myblue}\textbf{1.000} & \cellcolor{myblue}\textbf{1.000} & \cellcolor{myblue}\textbf{1.000} & \cellcolor{myblue}\textbf{1.000} \\
\midrule
\multirow{5}{*}{I2V} 
    & RivaGAN & 32 & 0.942 & 0.858 & 0.912 & 0.927 & 0.561 & 0.815 & 0.919 & 0.909 & 0.919 & 0.916 & 0.886 & 0.893 & \underline{0.781} & 0.853 \\
    & REVMark & 96 & 0.975 & 0.853 & 0.900 & 0.918 & 0.500 & 0.793 & 0.967 & - & - & - & 0.928 & 0.713 & 0.518 & 0.720 \\
    & VideoSeal & 96 & 0.982 & \underline{0.859} & \underline{0.915} & \underline{0.928} & \underline{0.573} & \underline{0.819} & 0.980 & \underline{0.980} & \underline{0.981} & \underline{0.981} & \underline{0.980} & \textbf{0.981} & 0.633 & 0.865 \\
    & VideoShield & 512 & \underline{0.990} & 0.811 & 0.892 & 0.913 & 0.530 & 0.787 & \underline{0.990} & - & - & - & \textbf{0.990} & 0.849 & 0.777 & \underline{0.872} \\
    & \cellcolor{myblue}\textit{VideoMark} & \cellcolor{myblue}{512$\times$15} & \cellcolor{myblue}\textbf{0.997} &
      \cellcolor{myblue}\textbf{0.864} & \cellcolor{myblue}\textbf{0.917} & \cellcolor{myblue}\textbf{0.930} & \cellcolor{myblue}\textbf{0.581} & \cellcolor{myblue}\textbf{0.823} &
      \cellcolor{myblue}\textbf{0.997} & \cellcolor{myblue}\textbf{0.991} & \cellcolor{myblue}\textbf{0.997} & \cellcolor{myblue}\textbf{0.995} &
      \cellcolor{myblue}0.857 & \cellcolor{myblue}\underline{0.955} & \cellcolor{myblue}\textbf{0.921} & \cellcolor{myblue}\textbf{0.911} \\
\bottomrule
\end{tabular}
\end{adjustbox}
\label{tab:main_result}
\end{table*}
%\end{comment}

%\input{tables/5_video_quality}

\subsubsection{Quality} The ``Video Quality'' columns report objective VBench scores. VideoMark obtains the highest average among the evaluated methods in both tasks. On I2V, its average exceeds VideoSeal by 0.004 and VideoShield by 0.036; its Image Quality scores are 0.692 on T2V and 0.581 on I2V.

Complementary GPT-4o evaluation also favors VideoMark in the tested settings. Appendix~\ref{MLLM_for_video_quality} reports the evaluation protocol, multi-method winner counts, mean scores, and watermark-free references.

\begin{table}[t]
    \centering
    % 第一个表格
    \begin{minipage}{0.48\textwidth}
        \centering
        \caption{Temporal-tampering detection p-values with $N=1000$ Monte Carlo references.}
        \label{tab:robustness_results_p_value}
        % 使用 adjustbox 限制宽度为当前 minipage 的宽度
        \begin{adjustbox}{width=\linewidth}
        \begin{tabular}{ccc|ccc}
            \toprule
            \multicolumn{3}{c}{\textbf{ModelScope}} & \multicolumn{3}{c}{\textbf{I2VGen-XL}} \\
            \cmidrule(l){1-3} \cmidrule(l){4-6}
            Swap & Insert & Drop & Swap & Insert & Drop \\
            \midrule
            0.001 & 0.001 & 0.001 & 0.001 & 0.001 & 0.001 \\
            \bottomrule
        \end{tabular}
        \end{adjustbox}
    \end{minipage}
    \hfill
    % 第二个表格
    \begin{minipage}{0.48\textwidth}
        \centering
        \caption{Comparison of temporal robustness, using matching accuracy as the evaluation metric.}
        \label{tab:robustness_results_frame_acc}
        % 重点在这里：强制第二个表缩放到 minipage 宽度
        \begin{adjustbox}{width=\linewidth}
        \begin{tabular}{cccc|ccc}
            \toprule
            \multirow{2}{*}{\textbf{Method}} & 
            \multicolumn{3}{c}{\textbf{ModelScope}} & 
            \multicolumn{3}{c}{\textbf{I2VGen-XL}} \\
            \cmidrule(l){2-7}
             & Swap & Insert & Drop & Swap & Insert & Drop \\
            \midrule
            VideoShield  & 1.000 & 1.000 & 1.000 & \underline{0.983} & \underline{0.983} & \underline{0.981} \\
            \textit{VideoMark} & \cellcolor{myblue}1.000 & \cellcolor{myblue}1.000 & \cellcolor{myblue}1.000 & \cellcolor{myblue}\textbf{0.996} & \cellcolor{myblue}\textbf{0.989} & \cellcolor{myblue}\textbf{0.996} \\
            \bottomrule
        \end{tabular}
        \end{adjustbox}
    \end{minipage}
\end{table}

\subsubsection{Robustness} The ``Temporal Tampering'' and ``Spatial Tampering'' columns report robustness under the attacks defined in Appendices~\ref{appendix:temporal_tampering} and~\ref{appendix:spatial_tampering}.

For temporal tampering, the main table reports bit accuracy between decoded and embedded messages. REVMark does not release the required model files. We omit VideoShield's insertion and drop entries from this table because its fixed-shape decoder does not produce a directly comparable payload after the frame count changes. The separate results in Table~\ref{tab:robustness_results_frame_acc} use a frame-matching metric under the controlled synthetic protocol and should not be interpreted as changed-length payload-decoding accuracy. VideoMark performs well on the synthetic swap, insertion, and drop operations in Table~\ref{tab:main_result}, but these attacks do not represent the broader range of video transformations encountered after distribution.

Table~\ref{tab:robustness_results_p_value} reports detection p-values under temporal manipulation; both models yield $p=0.001$ at the reported precision. Table~\ref{tab:robustness_results_frame_acc} compares frame-index matching with VideoShield. VideoMark reaches 0.996 on I2V, while the component ablation in Appendix~\ref{sec:component_ablation} isolates the contribution of TMM.

For spatial tampering, VideoMark embeds 32 bits per frame over 16 frames, for the same 512-bit total payload used by VideoShield. It achieves 1.000 average accuracy on T2V and 0.911 on I2V in this matched-total-payload setting. At the selected 512-bit-per-frame operating point, strong blur, resizing, and H.264 compression expose the expected capacity--robustness trade-off (Section~\ref{sec:payload_robustness}).

\subsubsection{Robustness under Video Editing}\label{sec:expanded_robustness}
Table~\ref{tab:realistic_temporal} compares message recovery under practical editing and compression on ModelScope. VideoMark uses 32 bits per frame, or 512 bits for the original 16-frame video, and VideoShield uses a 512-bit video payload. \rev{The VideoShield evaluation contains 100 videos per attack. For operations supported by both methods, VideoShield retains higher bit accuracy under block swapping and H.264 compression, while VideoMark performs better under cross-fading.} Appendix~\ref{sec:payload_robustness} reports additional VideoMark results at 512 bits per frame and under latent-space perturbations.

The editing operations affect watermark recovery in different ways. TMM recovers correspondence under temporal reordering. Trimming, speed changes, and frame-rate conversion remove or resample observations, while frame averaging, cross-fading, and compression alter or mix frame content. VideoMark supports extraction from the evaluated changed-length inputs; the resulting bit accuracies quantify how much message information survives each operation.

Appendix~\ref{sec:detection_reliability} complements these message-recovery results with video-level TPR, FPR, and threshold sensitivity.

\begin{table}[t]
\centering
\caption{Mean bit accuracy under video editing on ModelScope. The original 16-frame payload is 512 bits: 32 bits/frame for VideoMark and 512 bits/video for VideoShield. \rev{A dash denotes an unsupported changed-length input for the evaluated VideoShield decoder.} Values near 0.5 indicate chance-level message recovery.}
\label{tab:realistic_temporal}
\small
\begin{tabular}{lccc}
\toprule
Attack & Frames after attack & VideoMark & \rev{VideoShield} \\
\midrule
Identity & 16 & 1.0000 & \rev{1.0000} \\
Head trim (25\%) & 12 & 0.5017 & \rev{--} \\
Middle trim (25\%) & 12 & 0.7466 & \rev{--} \\
Tail trim (25\%) & 12 & 0.9926 & \rev{--} \\
Playback speed ($0.5\times$) & 32 & 0.5300 & \rev{--} \\
Playback speed ($2\times$) & 8 & 0.5950 & \rev{--} \\
FPS conversion (10$\rightarrow$24) & 38 & 0.5068 & \rev{--} \\
FPS conversion (10$\rightarrow$30) & 48 & 0.5083 & \rev{--} \\
Block swap & 16 & 0.9899 & \rev{1.0000} \\
Frame averaging & 16 & 0.5013 & \rev{0.5223} \\
Cross-fade & 16 & 0.6583 & \rev{0.5801} \\
H.264 (CRF18) & 16 & 0.6867 & \rev{1.0000} \\
H.264 (CRF23) & 16 & 0.6348 & \rev{0.9995} \\
H.264 (CRF28) & 16 & 0.5220 & \rev{0.9901} \\
\bottomrule
\end{tabular}
\end{table}

\subsubsection{Detectability} 
\begin{figure}[!htbp]
    \centering
    \begin{minipage}{0.49\textwidth}
        \centering
        \includegraphics[width=\linewidth]{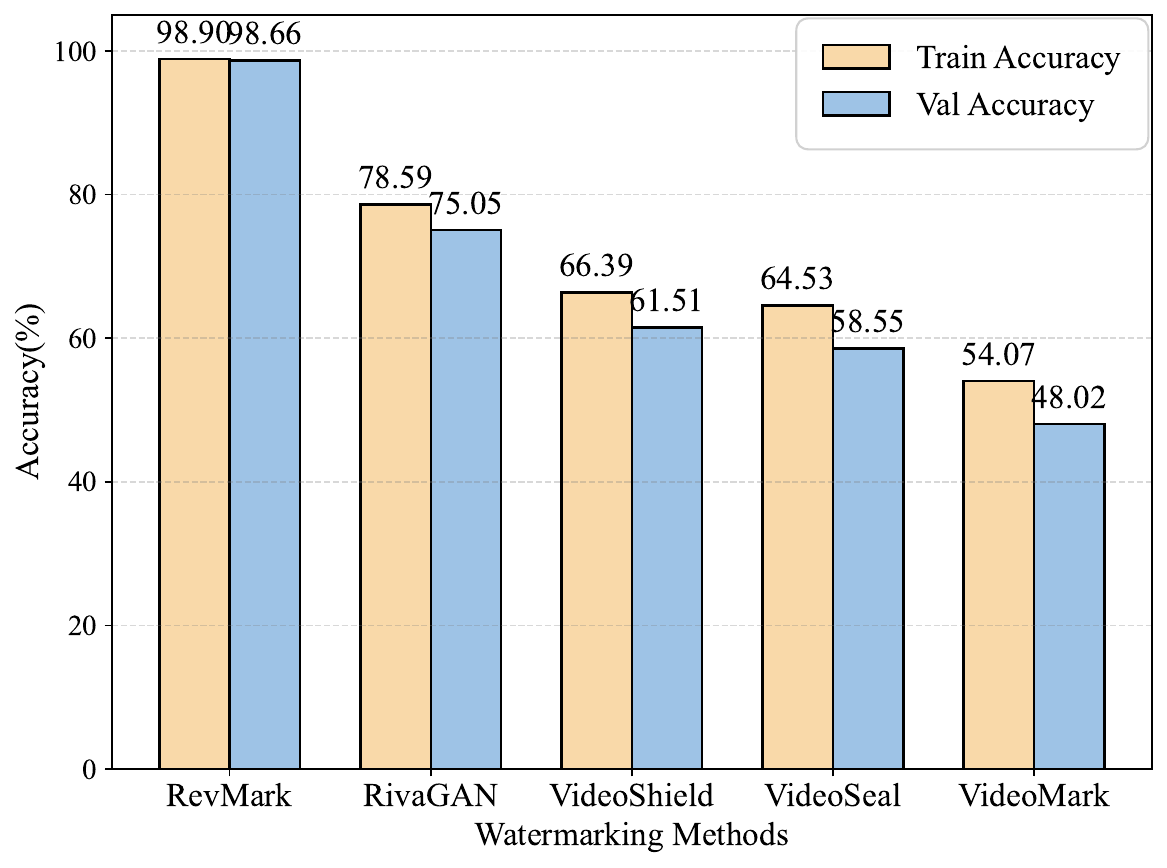}
        \caption{The binary classification results under different watermarking algorithms.}
        \label{fig:classfication}
    \end{minipage}
    \hfill
    \begin{minipage}{0.49\textwidth}
        \centering
        \includegraphics[width=\linewidth]{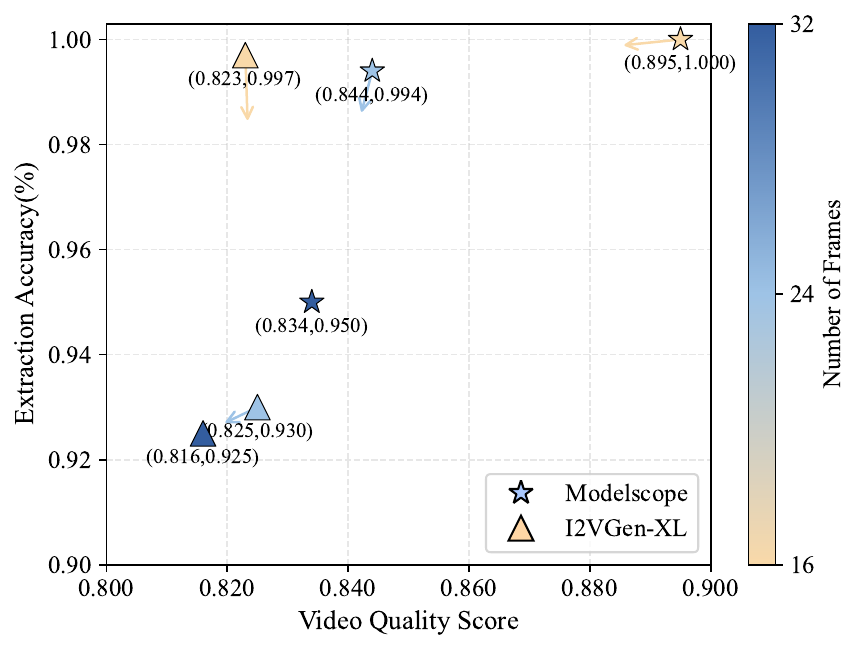}
        \caption{The extraction accuracy and visual quality score of the watermarking scheme across different models and frame counts.} 
        \label{fig:ablation_frame_length}
    \end{minipage}
\end{figure}

We train a VideoMAE classifier~\citep{tong2022videomae} for 100 epochs on 8,000 watermark-free and 8,000 watermarked videos. Figure~\ref{fig:classfication} shows that VideoMark classification remains close to chance, reaching 54.07\% training accuracy and 48.02\% validation accuracy. The larger train--validation agreement of the other methods indicates that their watermark patterns are easier for this classifier to learn. This experiment evaluates one learned detector and should not be interpreted as proof against every adaptive detector.

\FloatBarrier

\subsection{Parameter Sensitivity}
\subsubsection{Impact of Video Length}\label{sec:long_video}
Figures~\ref{fig:ablation_frame_length} and~\ref{fig:frame_quality} examine extraction accuracy and generation quality as frame count increases. Figure~\ref{fig:ablation_frame_length} shows that the main effect of longer sequences is a reduction in extraction accuracy. In Figure~\ref{fig:frame_quality}, watermarked and watermark-free quality distributions substantially overlap at 16, 24, and 32 frames for both backbones. Their absolute quality distributions vary with the generation model and video length, while the watermark preserves comparable visual quality at each tested length. Together, these results characterize how generation quality and watermark recovery vary with sequence length. Appendix~\ref{sec:parameter_sensitivity} reports the extended length evaluation and configuration analyses.

\subsubsection{Impact of Message Length}
\begin{figure}[!htbp]
    \centering
    \begin{minipage}[b]{0.48\textwidth}
        \centering
        \includegraphics[width=\linewidth]{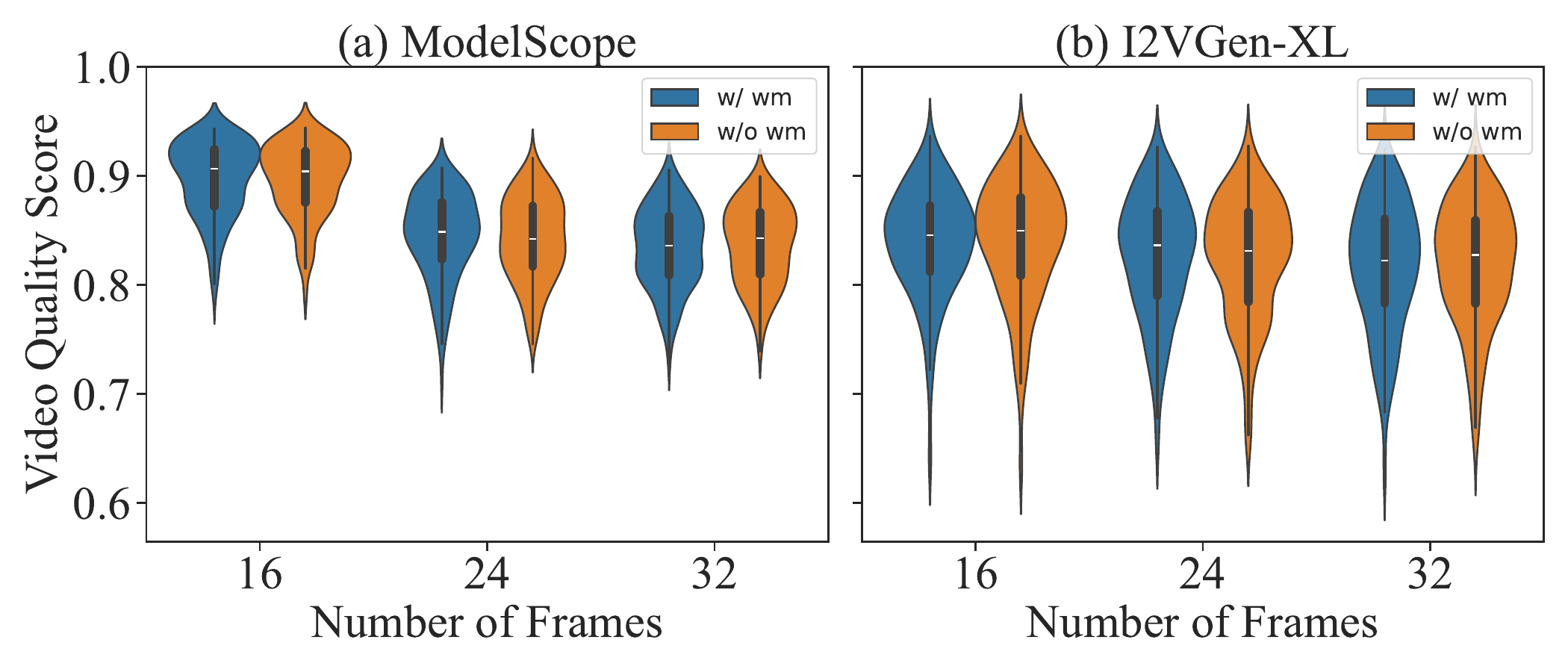}
        \caption{Video quality scores under varying numbers of frames, comparing watermarked (w/ wm) and clean (w/o wm) outputs.}
        \label{fig:frame_quality}
    \end{minipage}
    \hfill
    \begin{minipage}[b]{0.48\textwidth}
        \centering
        \includegraphics[width=\linewidth]{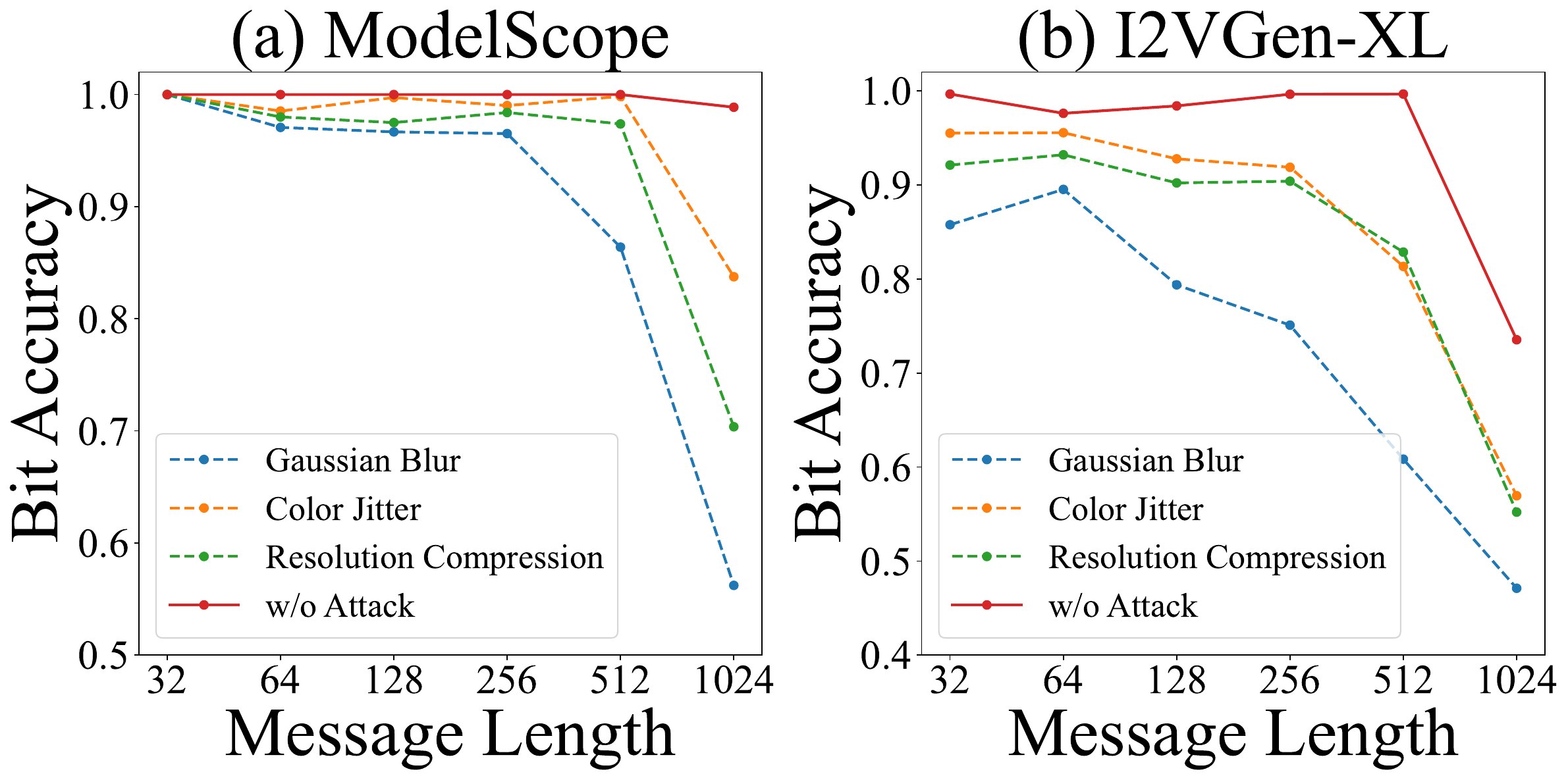}
        \caption{The extraction accuracy and robustness of VideoMark against spatial tampering for varying message lengths.}
        \label{fig:ablation_message}
    \end{minipage}
\end{figure}

Figure~\ref{fig:ablation_message} evaluates clean extraction and three spatial transformations across payload sizes. Clean accuracy remains near saturation through 512 bits per frame and drops at 1024 bits. Robustness changes earlier: Gaussian blur becomes more damaging as payload grows, while color jitter has a smaller effect at low payloads and reduces accuracy at the largest settings. Resolution compression produces the lowest recovery accuracy across most settings, particularly for I2VGen-XL. The backbones also exhibit different payload responses: ModelScope's robustness is highest at 32 bits, whereas some I2VGen-XL attack curves peak or remain comparable around 64 bits. These results motivate using up to 512 bits per frame for clean extraction and the matched 32-bit-per-frame setting for robustness comparisons. Appendix~\ref{sec:payload_robustness} provides numerical results across payloads and attack strengths.

\begin{comment}
\begin{figure}[!tbp]
    \centering
    \includegraphics[width=0.85\linewidth]{figures/5_quality_distribution.pdf}
    \caption{Video quality scores under varying numbers of frames, comparing watermarked (w/ wm) and clean (w/o wm) outputs.}
    \label{fig:frame_quality}
\end{figure}
\begin{figure}[!tbp]
    \centering
    \includegraphics[width=0.85\textwidth]{figures/5_ablation_message.pdf}
    \caption{The extraction accuracy and robustness of VideoMark against spatial tampering for varying message lengths.}
    \label{fig:ablation_message}
\end{figure}
\end{comment}

\FloatBarrier

\section{Scope and Practical Considerations}
\label{sec:scope_limitations}

\textbf{Robustness and verification outcomes.} Temporal matching recovers correspondence when frame messages remain decodable. Table~\ref{tab:realistic_temporal} and Appendix~\ref{sec:payload_robustness} characterize recovery under reordering, removal, resampling, and content-altering attacks. Detection, bit recovery, and temporal recovery measure complementary verification outcomes. \rev{Unsuccessful detection does not establish that a video was not AI-generated.}

\textbf{Verification setup and efficiency.} Verification requires the corresponding model/VAE, scheduler, preprocessing and inversion settings, private PRC key, and registered sequence. \rev{Original prompts and conditioning images are unnecessary: T2V uses an empty prompt; I2V uses the literal string \texttt{\char34 none\char34} and the observed first frame.} This setup suits provider-managed or authorized verification services. For 16 frames and 25 DDIM steps, extraction takes $26.40\pm1.09$ seconds and 10.17GB peak GPU memory; per-frame BP-OSD takes $1.626\pm0.060$ seconds. Inversion and decoding dominate matching cost (Appendix~\ref{sec:runtime}), favoring offline use.

\textbf{Architecture and attribution scope.} Evaluation covers frame-aligned 2D VAEs and recovery of registered watermark messages. Extending the design to temporally compressed causal-3D-VAEs requires group-level embedding and alignment, as illustrated by SIGMark~\citep{zhu2026sigmark}. A multi-user attribution service additionally maps recovered messages to users or videos through its registration protocol. \rev{Deployment requires protected keys, a registration/search protocol, calibrated decisions, and review of disputed matches.}

\section{Conclusion}
\label{conclusion}

VideoMark combines frame-wise PRC initialization, random-start registration, and temporal matching for training-free watermarking of frame-aligned 2D-VAE video diffusion models. \rev{Under the encoder's joint pseudorandomness assumption, the construction provides a joint distribution guarantee under key reuse.} Experiments on ModelScope and I2VGen-XL show accurate clean extraction and quality comparable to watermark-free generation. Temporal matching improves recovery under reordering, \rev{and the editing experiments characterize watermark detection and message recovery across temporal transformations and compression.} Future work includes faster and blind verification, greater resilience to temporal resampling, and extensions to temporally compressed causal-3D-VAEs.

\bibliography{main}
\bibliographystyle{tmlr}

\appendix
%\section{Appendix}
%You may include other additional sections here.
\clearpage
\section{Appendix}
\raggedbottom
\subsection{PRC Watermark}
\label{prc_watermark}
% Source: October 21 rebuttal to qJui (v4dWu4eD2Z, S7h7VmhViW)
% and K35z (raycf8EWSl). This is a conditional reduction, not a new PRC proof.
\begin{revision}
\subsubsection{Conditional Joint Distribution Guarantee}
\label{sec:joint_prc_guarantee}
We distinguish exact Gaussian equality for ideal independent uniform signs from the computational guarantee inherited by the implemented PRC construction. The latter concerns the entire initialization tensor, including multiple frames encoded under one secret key.

\paragraph{Encoding model.}
Let $f$ be the number of latent frames and $n=chw$ the number of embedded coordinates per frame. Write the binary PRC codeword as
\begin{equation}
 C_i=A v_i\oplus\boldsymbol{b}\oplus e_i,
 \qquad v_i=s\mathbin\Vert r_i\mathbin\Vert m_i\mathbin\Vert 0,
 \qquad i=1,\ldots,f.
\end{equation}
Here $A$ is the binary generator matrix, $\boldsymbol{b}$ is the key-dependent mask, $s$ contains test bits, $m_i$ is the frame message, $r_i$ is a fresh random prefix, $e_i$ is fresh encoding noise, and the final zero vector provides padding if needed. Arithmetic is over $\mathbb{F}_2$. The same key is reused across frames, with fresh randomness for each encoding invocation. In the sign notation of Section~\ref{subsec:generation}, $\mathbf{c}_i=1-2C_i$. Independently sample $G_i\sim\mathcal{N}(0,I_n)$ and form
\begin{equation}
 Z_i=(1-2C_i)\odot|G_i|.
\end{equation}

\paragraph{Assumption: joint pseudorandomness.}
Let $K\leftarrow\operatorname{KeyGen}(1^\lambda)$ be sampled once and kept secret. For the message-generation process and parameter family under consideration, assume that every efficient distinguisher $B$ satisfies
\begin{equation}
 \left|\Pr[B(C_1,\ldots,C_f)=1]-\Pr[B(U_1,\ldots,U_f)=1]\right|
 \leq\varepsilon_{\mathrm{PRC}}(\lambda,f,B),
 \label{eq:joint_prc_assumption}
\end{equation}
where $U_1,\ldots,U_f$ are independent uniform elements of $\{0,1\}^n$. Probabilities include key generation and encoding randomness; $B$ does not receive the secret key. Any public conditioning or auxiliary information must be included consistently in both experiments and covered by this assumption. Individual-codeword pseudorandomness and fresh encoding randomness alone do not establish this joint property under key reuse.

\paragraph{Proposition and proof.}
Under Eq.~\eqref{eq:joint_prc_assumption}, for every efficient latent distinguisher $D$ there is an efficient codeword distinguisher $B_D$ such that
\begin{equation}
 \left|\Pr[D(Z_1,\ldots,Z_f)=1]-\Pr[D(H_1,\ldots,H_f)=1]\right|
 \leq\varepsilon_{\mathrm{PRC}}(\lambda,f,B_D),
 \label{eq:joint_latent_guarantee}
\end{equation}
where the $H_i$ are independent $\mathcal{N}(0,I_n)$ samples. Define the randomized mapping
\begin{equation}
 \Phi(x;g)=\big((1-2x_i)\odot|g_i|\big)_{i=1}^{f}.
\end{equation}
For uniform $U$, each coordinate has an independent uniform sign and an independent half-normal magnitude. If $\varphi$ is the standard normal density, the half-normal density at $y>0$ is $2\varphi(y)$; selecting either sign with probability $1/2$ gives density $\varphi$ on both sides of zero. Independence across coordinates and frames therefore implies the joint identity
\begin{equation}
 \Phi(U;G)\sim\mathcal{N}(0,I_{nf}).
\end{equation}
Given a codeword tuple $x$, the reduction $B_D$ independently samples $G$ and returns $D(\Phi(x;G))$. PRC inputs reproduce the watermarked latent experiment, while uniform inputs reproduce the Gaussian reference. Thus
\begin{equation}
 \operatorname{Adv}_{\mathrm{latent}}(D)
 =\operatorname{Adv}_{\mathrm{PRC}}(B_D),
\end{equation}
which proves Eq.~\eqref{eq:joint_latent_guarantee}. If the assumed PRC advantage is negligible for the chosen parameter family and a polynomial number of frames, the joint watermarked tensor is computationally indistinguishable from the Gaussian reference.

\paragraph{Extension through video generation.}
Let $F_\theta$ be the same efficient video generator in both experiments, with the same public conditioning $a$ and additional randomness $\xi,\xi'$ drawn from the same distribution independently of the latents:
\begin{equation}
 V_{\mathrm{wm}}=F_\theta(a,Z;\xi),
 \qquad V_0=F_\theta(a,H;\xi').
\end{equation}
A video distinguisher yields a latent distinguisher by first applying $F_\theta$. Composing this procedure with the preceding reduction bounds its advantage by the corresponding $\varepsilon_{\mathrm{PRC}}$. The generator may contain temporal interactions; generated frames need not be independent.

\paragraph{Scope of the guarantee.}
This reduction states how VideoMark inherits the encoder's joint pseudorandomness; it does not independently establish a negligible advantage for every concrete PRC parameter choice. The theoretical security conditions and practical parameters of PRC-Watermark~\citep{gunn2025an} must remain distinguished. Exact Gaussian equality applies to the ideal uniform-sign reference, not to every fixed PRC key. With finite-precision sampling, the same reduction applies relative to the uniform-sign reference using the same magnitude sampler and post-processing. The learned-detector experiment is complementary evidence for the tested implementation, not a substitute for the joint pseudorandomness assumption.
\end{revision}

\subsubsection{Detection Phase} \textbf{PRC.Detect} accepts as input a vector $\mathbf{s}$ with entries in the interval $[-1, 1]$ rather than a bit-string. \rev{For a binary codeword coordinate $C_j\in\{0,1\}$, $s_j$ is intended to approximate the conditional expectation of its sign $1-2C_j$ given the observation; this sign corresponds to the $\{-1,1\}$ codeword notation in the main text.} The full procedure for \textbf{PRC.Detect} is presented in Algorithm \ref{alg:prc_detect}.

\begin{algorithm}[H]
    \caption{PRC.Detect}
    \label{alg:prc_detect}
    \SetKwInOut{Input}{Input}
    \SetKwInOut{Output}{Output}
    
    \Input{PRC-key $k$, $s$}
    \Output{Detection result True or False}
    \BlankLine

    Parse parameters from key $k$: $(n, \text{msg\_len}, F, t, \lambda, \eta, \text{bits}, k, r, \text{iter}, \text{otp}, \text{testbits}, G, P) \gets k$\;
    
    \For{$i \in [n]$}{
        $s_i \gets (-1)^{\text{otp}_i} \cdot (1 - 2\eta) \cdot s_i$\;
    }
    
    \For{each parity check $w \in P$}{
        Let $\hat{s}_w \gets \prod_{i \in w} s_i$\;
    }

    $C \gets \frac{1}{2} \sum_{w \in P} \log^2\left(\frac{1+\hat{s}_w}{1-\hat{s}_w}\right)$\;

    \eIf{$\sum_{w \in P} \log\left(\frac{1 + \hat{s}_w}{2}\right) \ge \sqrt{C \log(1/F)} + \frac{1}{2} \sum_{w \in P} \log\left(\frac{1 - \hat{s}_w^2}{4}\right)$}{
        \Return True\;
    }{
        \Return False\;
    }
\end{algorithm}

\subsubsection{Decoding Phase} \textbf{PRC.Decode} accepts as input a vector $\mathbf{s}$ with entries in the interval $[-1, 1]$, and returns a decoded binary message. The full procedure for \textbf{PRC.Decode} is presented in Algorithm~\ref{alg:prc_decode}.

\begin{algorithm}[H]
    \caption{PRC.Decode}
    \label{alg:prc_decode}
    \SetKwInOut{Input}{Input}
    \SetKwInOut{Output}{Output}
    
    \Input{PRC-key $k$, $s$}
    \Output{Decoded message $m \in \{0,1\}^k$ or $\text{None}$}
    \BlankLine

    Parse parameters from key $k$: $(n, \text{msg\_len}, F, t, \lambda, \eta, \text{bits}, k, r, \text{iter}, \text{otp}, \text{testbits}, G, P) \gets k$\;
    
    \For{$i \in [n]$}{
        $s_i \gets (-1)^{\text{otp}_i} \cdot (1 - 2\eta) \cdot s_i$\;
    }
    
    $y \gets \text{BP-OSD}(G, P, s)$\;
    Parse $( \text{testbits}, r, m ) \gets y$\;
    
    \Return $m$\;
\end{algorithm}

Algorithm~\ref{alg:prc_detect} tests whether a frame contains a valid PRC codeword without recovering its payload. Algorithm~\ref{alg:prc_decode} additionally recovers the binary frame message used by TMM.

\begin{table*}[h]
	\centering
    \caption{Video quality comparison across watermarking methods. VideoMark achieves the highest or near-highest scores across all metrics, closely matching clean videos.
    }
    \resizebox{\textwidth}{!}{
	\begin{tabular}{c|lcccc}
		\midrule
		  Model & Method & Subject Consistency & Background Consistency &  Motion Smoothness  & Imaging Quality\\
		 
	  	\midrule
		\multirow{6}*{MS} & RivaGAN & 0.922 & 0.951 &  0.960  & 0.648\\
		
		& REVMark & 0.943 & 0.960 &  0.972  & 0.450\\
	    
        & VideoSeal & 0.950  &  0.959 &  \textbf{0.977}  & 0.679\\

		& VideoShield &0.949  & \textbf{0.962}  & \textbf{0.977} & 0.689 \\
		
        & \cellcolor{myblue} \textit{VideoMark (ours)} & \cellcolor{myblue}\textbf{0.951} & \cellcolor{myblue}0.961 &  \cellcolor{myblue}\textbf{0.977}  & \cellcolor{myblue}\textbf{0.692}\\
		
        & w/o watermark & \textbf{0.951} & \textbf{0.962}  &\textbf{0.977}  & \underline{0.691} \\
		\midrule
        		
		\multirow{6}*{I2V} & RivaGAN & 0.858 & 0.912 &  0.927 & 0.561\\

            & REVMark & 0.853 & 0.900 &  0.918  & 0.500 \\
            
            & VideoSeal & 0.859  & \underline{0.915} & \underline{0.928}  & 0.573 \\
    
            & VideoShield & 0.811  &0.892  &0.913 & 0.530   \\
            
		  & \cellcolor{myblue} \textit{VideoMark (ours)} & \cellcolor{myblue}\textbf{0.864} & \cellcolor{myblue}\textbf{0.917} &  \cellcolor{myblue}\textbf{0.930} & \cellcolor{myblue}\textbf{0.581}\\
		  
            & w/o watermark & \underline{0.861} & 0.913 &  \underline{0.928} & \underline{0.578}\\

            \bottomrule
	\end{tabular}
    }
	\label{tab:support video quality}
\end{table*}
\subsection{VBench Video-Quality Metrics}
\label{vBench_for_Video_Quality}
We summarize the four VBench metrics used in our evaluation: Subject Consistency, Background Consistency, Motion Smoothness, and Imaging Quality.
\begin{itemize}
    \item \textbf{Subject Consistency} is measured by computing the similarity of DINO features~\citep{caron2021emerging} across frames:
\begin{equation}
    S_{\text{subject}} = \frac{1}{T - 1} \sum_{t=2}^{T} \frac{1}{2} \left( \langle d_1, d_t \rangle + \langle d_{t-1}, d_t \rangle \right),
\end{equation}
where $d_i$ denotes the DINO image feature of the $i^\text{th}$ frame, normalized to unit length, and $\langle \cdot, \cdot \rangle$ represents the dot product, corresponding to cosine similarity.

\item \textbf{Background Consistency} assesses the temporal coherence of background scenes by computing CLIP feature similarity across frames \citep{radford2021learning}:
\begin{equation}
    S_{\text{background}} = \frac{1}{T - 1} \sum_{t=2}^{T} \frac{1}{2} \left( \langle c_1, c_t \rangle + \langle c_{t-1}, c_t \rangle \right),
\end{equation}
where $c_i$ denotes the CLIP image feature of the $i^\text{th}$ frame, normalized to unit length.

\item \textbf{Motion Smoothness} is assessed using a frame-by-frame motion prior inspired by video frame interpolation models \citep{li2023amt}.
Given a generated video with frames \([f_0, f_1, f_2, f_3, f_4, \ldots, f_{2n-2}, f_{2n-1}, f_{2n}]\), the odd-indexed frames are manually removed to create a low-frame-rate sequence \([f_0, f_2, f_4, \ldots, f_{2n-2}, f_{2n}]\). A video frame interpolation model \citep{li2023amt} is then applied to reconstruct the dropped frames \([\hat{f}_1, \hat{f}_3, \ldots, \hat{f}_{2n-1}]\).
The Mean Absolute Error (MAE) between the reconstructed and original frames is calculated and normalized to the range \([0, 1]\), where a higher value indicates smoother motion.

\item \textbf{Imaging Quality} is evaluated using the MUSIQ image quality predictor \citep{ke2021musiq}, which is trained on the SPAQ dataset \citep{fang2020perceptual} and is capable of handling images with varying aspect ratios and resolutions.
Each frame is assigned a quality score by MUSIQ, which is linearly normalized to the range \([0, 1]\) by dividing by 100. The final imaging quality score is computed by averaging the normalized scores across all frames in the video.
\end{itemize}

Table~\ref{tab:support video quality} compares watermarked and watermark-free video quality on VBench. VideoMark preserves quality comparable to watermark-free generation.

\subsection{Qualitative Video Results}
\label{sec:qualitative_results}
\begin{figure*}[t]
  \centering
  \includegraphics[width=\linewidth]{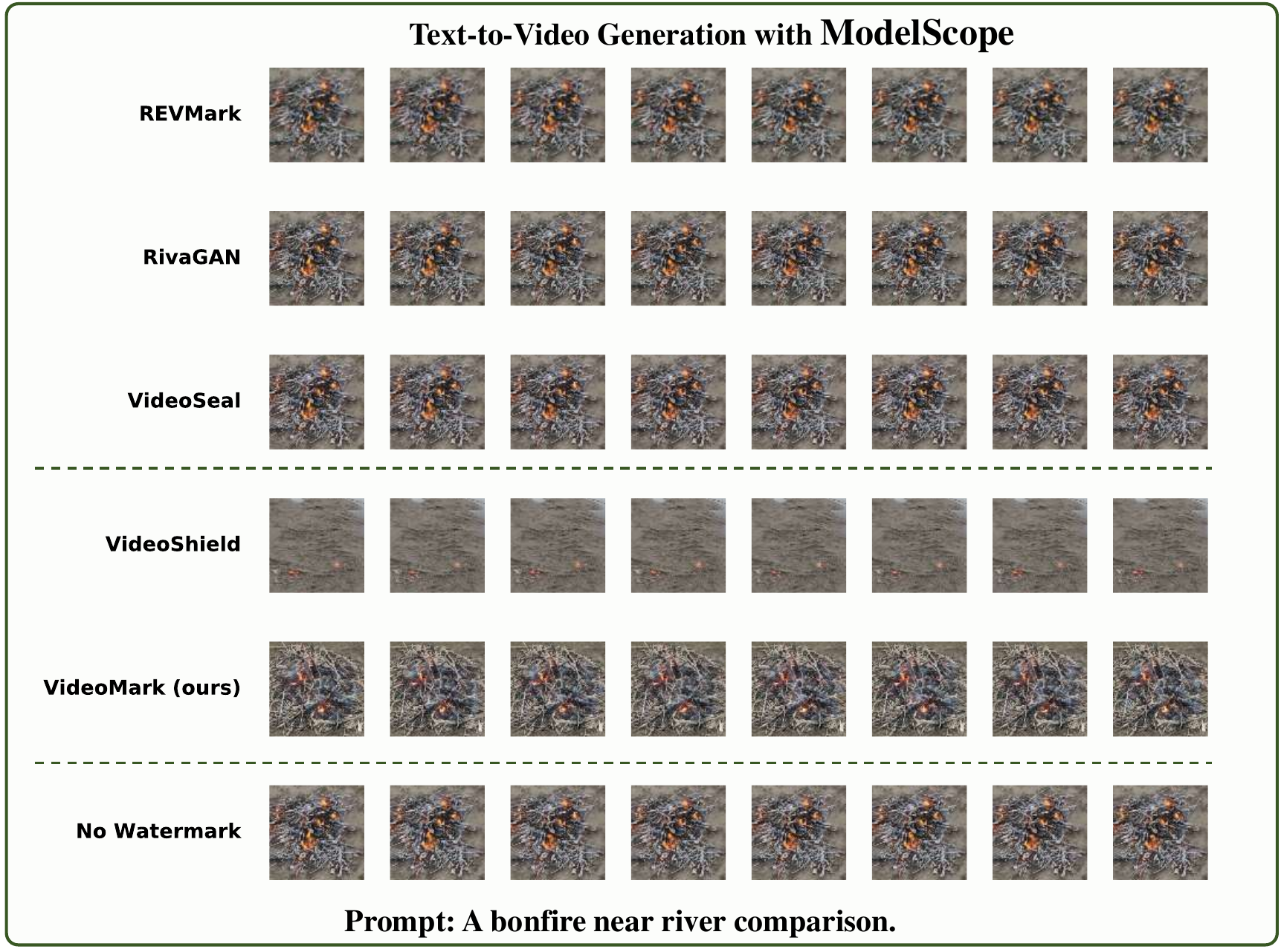}
  \caption{Visualization results comparing VideoMark with existing approaches on the T2V task.}
  \label{fig:visualization_t2v}
\end{figure*}

\begin{figure*}[t]
  \centering
  \includegraphics[width=\linewidth]{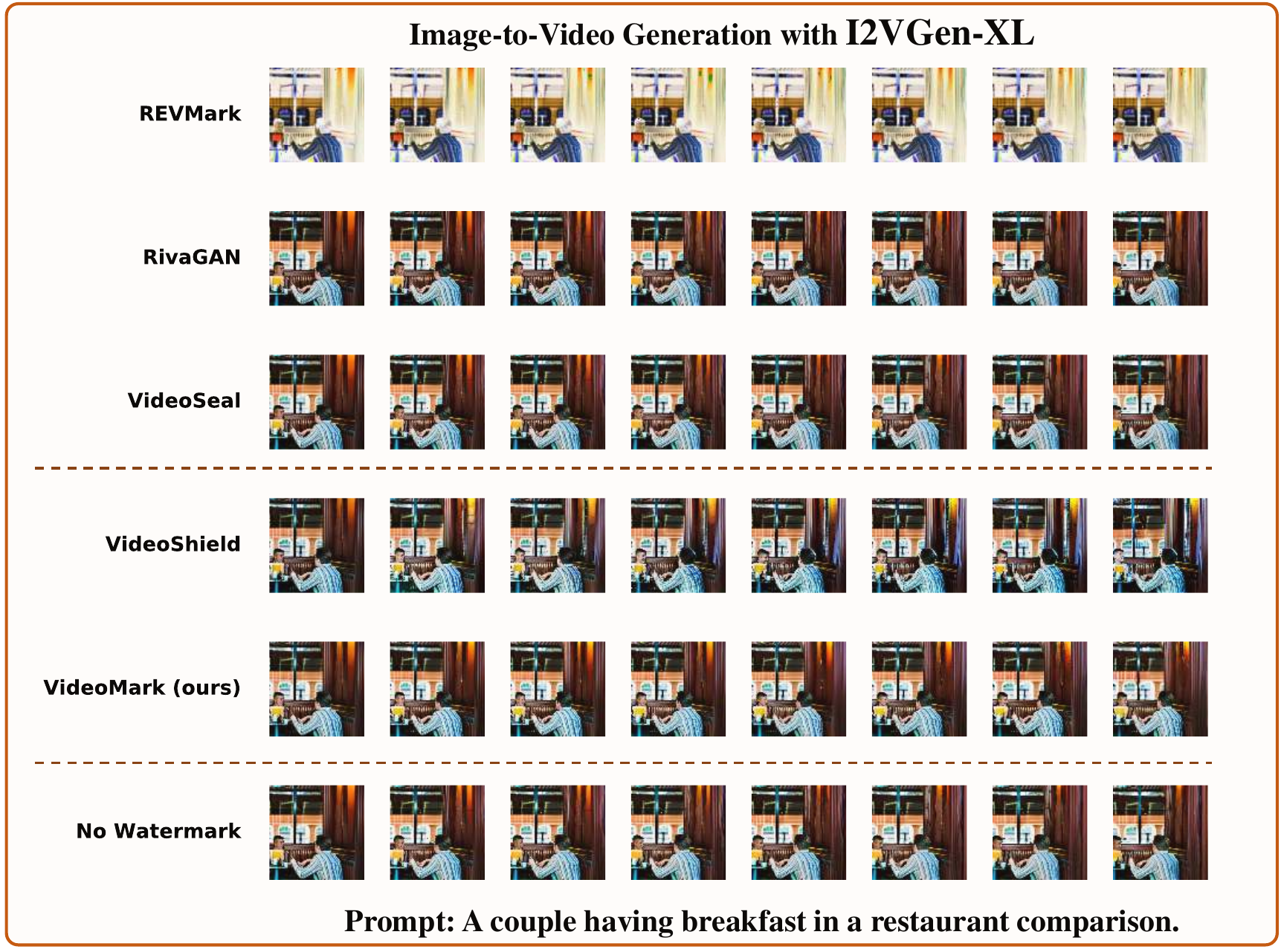}
  \caption{Visualization results comparing VideoMark with existing approaches on the I2V task.}
  \label{fig:visualization_i2v}
\end{figure*}

Figures~\ref{fig:visualization_t2v} and~\ref{fig:visualization_i2v} show the first eight frames of representative T2V and I2V samples. Across both tasks, VideoMark remains visually close to watermark-free generation and avoids the conspicuous degradation observed for some baselines. These examples are qualitative and complement, rather than replace, the aggregate VBench and GPT-4o evaluations.

\begin{figure*}[htbp] 
\centering
\begin{tcolorbox}[
    colback=myblue!5!white,
    colframe=black!50!myblue,
    coltitle=black,
    %boxsep=6pt,
    colbacktitle=myblue!20!white,
    fonttitle=\bfseries,
    title=Prompts for GPT-4o Scoring Video Quality,
]
%\small
Please perform a thorough and objective evaluation of the video result for sample \texttt{'sample\_name'}. Using the complete ordered sequence of all 16 video frames, carefully assess the video on the following criteria:\\[1mm]

\textbf{1. Subject Consistency:} Evaluate how consistently and accurately the subject is represented across all frames.\\
\textbf{2. Background Consistency:} Assess the uniformity and coherence of the background elements throughout the video.\\
\textbf{3. Motion Smoothness:} Judge the fluidity and natural progression of motion, ensuring transitions are smooth and realistic.\\
\textbf{4. Imaging Quality:} Evaluate the overall visual quality, including clarity, sharpness, color fidelity, and detail.\\[1mm]

For each criterion, assign a numeric score between 0 (lowest) and 10 (highest) based solely on the visual data. Compute the total score as the sum of these individual scores. Your evaluation should be objective, data-driven, and free from any additional commentary or explanation. Return your response strictly as a JSON object formatted exactly as shown below without any extra text:\\[1mm]

\texttt{\{ }\\
\quad \texttt{"subject\_consistency": \textless number\textgreater,}\\
\quad \texttt{"background\_consistency": \textless number\textgreater,}\\
\quad \texttt{"motion\_smoothness": \textless number\textgreater,}\\
\quad \texttt{"imaging\_quality": \textless number\textgreater,}\\
\quad \texttt{"total\_score": \textless number\textgreater }\\
\texttt{\}}\\[1mm]

\end{tcolorbox}
\caption{Prompt used for the subjective evaluation of video quality.}
\label{fig:prompt_for_evaluating_wm_videos}
\end{figure*}

\subsection{MLLM-as-a-Judge Evaluation}
\label{MLLM_for_video_quality}
As a complementary perceptual evaluation, GPT-4o scores the complete ordered frame sequences. We randomly sample 1,000 of the 8,000 videos generated by each model. Table~\ref{tab:gpt_quality} reports the number of samples on which each method receives the highest overall score in the multi-method evaluation. VideoMark ranks first on 288 T2V and 245 I2V samples, 48 and \rev{23} more than the respective runners-up. GPT-4o mean total scores are 28.32/28.65 for VideoMark and 27.85/28.36 for VideoShield on T2V/I2V, respectively, compared with watermark-free references of 28.31/28.73. These results complement the VBench evaluation of video quality.

\begin{table}[t]
\centering
\caption{Quality Assessment via GPT-4o. We report the total number of top-rated samples assigned by GPT-4o across different base models. The methods compared include RivaGAN (RG), REVMark (RM), VideoSeal (VS), VideoShield (VSh), and \textit{VideoMark} (\textit{VM}).}
\label{tab:gpt_quality}
\begin{tabular}{l|ccccc}
    \toprule
    \textbf{Model} & RG & RM & VS & VSh & \textit{VM} \\
    \midrule
    ModelScope & 231 & 47 & 194 & \underline{240} & \cellcolor{myblue}\textbf{288} \\
    I2VGen-XL  & \underline{222} & 111 & 205 & 217 & \cellcolor{myblue}\textbf{245} \\
    \midrule
    \textbf{Total} & 453 & 158 & 399 & \underline{457} & \cellcolor{myblue}\textbf{533} \\
    \bottomrule
\end{tabular}
\end{table}
%\begin{table*}[htbp]
%\centering
%\caption{Detailed performance evaluation of videos assessed by GPT-4o.}
%\label{tab:model_comparison}
%\resizebox{\textwidth}{!}{%
%\begin{tabular}{l c c c c c  c c c c c}
%\toprule
%\multirow{2}{*}{\textbf{Method}} & \multicolumn{4}{c}{\textbf{ModelScope}} & \multirow{2}{*}{Total} & \multicolumn{4}{c}{\textbf{I2VGen-XL}} & \multirow{2}{*}{Total} \\
%\cmidrule(lr){2-5} \cmidrule(lr){7-10}
%  & SC & BC & MS & IQ &  & SC & BC & MS & IQ &  \\
%\midrule
%RivaGAN     & \textbf{7.88} & 8.85 & \textbf{4.75} & 6.90 & \textbf{28.38} & \textbf{7.69} & 7.67 & \textbf{6.32} & 7.14 & \textbf{28.82}  \\
%REVMark     & 6.52 & 7.95 & 4.41 & 3.95 & 22.83 & 7.62 & \textbf{7.73} & 5.94 & 5.57 & 26.86  \\
%VideoSeal   & 7.82 & \underline{8.86} & 4.65 & 6.90 & 28.23 & 7.63 & 7.62 & 6.19 & 7.10 & 28.54  \\
%VideoShield & 7.77 & \underline{8.86} & 4.37 & 6.85 & 27.85 & 7.58 & 7.57 & 6.17 & 7.04 & 28.36  \\
%\rowcolor{myblue}
%\textit{VideoMark (ours)}   & \underline{7.86} & \textbf{8.93} & 4.55 & \textbf{6.98} & \underline{28.32} & 7.66 & \underline{7.70} & 6.10 & \textbf{7.19} & 28.65  \\
%w/o watermark & 7.84 & 8.84 & \underline{4.67} & \underline{6.95} & 28.31 & \underline{7.67} & 7.61 & \underline{6.28} & \underline{7.17} & \underline{28.73} \\
%\bottomrule
%\end{tabular}%
%}
%\begin{minipage}{\textwidth}
%\centering
%\footnotesize
%SC: Subject Consistency \hspace{1.5em} 
%BC: Background Consistency \hspace{1.5em} 
%MS: Motion Smoothness \hspace{1.5em} \\
%IQ: Imaging Quality
%\end{minipage}
%\end{table*}

\begin{table*}[htbp]
\centering
\caption{Detailed performance evaluation of videos assessed by GPT-4o. Metrics include SC (Subject Consistency), BC (Background Consistency), MS (Motion Smoothness), and IQ (Imaging Quality).}
\label{tab:model_comparison}
%\resizebox{\textwidth}{!}{%
\begin{tabular}{l c c c c c  c c c c c}
\toprule
\multirow{2}{*}{\textbf{Method}} & \multicolumn{4}{c}{\textbf{ModelScope}} & \multirow{2}{*}{Total} & \multicolumn{4}{c}{\textbf{I2VGen-XL}} & \multirow{2}{*}{Total} \\
\cmidrule(lr){2-5} \cmidrule(lr){7-10}
  & SC & BC & MS & IQ &  & SC & BC & MS & IQ &  \\
\midrule
RivaGAN     & \textbf{7.88} & 8.85 & \textbf{4.75} & 6.90 & \textbf{28.38} & \textbf{7.69} & 7.67 & \textbf{6.32} & 7.14 & \textbf{28.82}  \\
REVMark     & 6.52 & 7.95 & 4.41 & 3.95 & 22.83 & 7.62 & \textbf{7.73} & 5.94 & 5.57 & 26.86  \\
VideoSeal   & 7.82 & \underline{8.86} & 4.65 & 6.90 & 28.23 & 7.63 & 7.62 & 6.19 & 7.10 & 28.54  \\
VideoShield & 7.77 & \underline{8.86} & 4.37 & 6.85 & 27.85 & 7.58 & 7.57 & 6.17 & 7.04 & 28.36  \\
\rowcolor{myblue}
\textit{VideoMark (ours)}   & \underline{7.86} & \textbf{8.93} & 4.55 & \textbf{6.98} & \underline{28.32} & 7.66 & \underline{7.70} & 6.10 & \textbf{7.19} & 28.65  \\
w/o watermark & 7.84 & 8.84 & \underline{4.67} & \underline{6.95} & 28.31 & \underline{7.67} & 7.61 & \underline{6.28} & \underline{7.17} & \underline{28.73} \\
\bottomrule
\end{tabular}%
%}
\end{table*}

Figure~\ref{fig:prompt_for_evaluating_wm_videos} shows the GPT-4o evaluation prompt, and Table~\ref{tab:model_comparison} reports mean scores. The model receives the complete frame sequence in temporal order, preserving the temporal context emphasized in continuous video-understanding benchmarks such as RTV-Bench~\citep{xun2025rtv}. The prompt asks the judge to score the observed visual data; related visual-grounding evaluations examine how structured prompting helps MLLMs attend to visual evidence~\citep{ling2026seeing}. VideoMark is competitive across the four criteria, although RivaGAN scores slightly higher on subject consistency and motion smoothness. Because MLLM judgments may be sensitive to prompting and model version, we use them only as a complement to the objective VBench metrics.

%\clearpage
\subsection{Temporal Tampering Implementation}
\label{appendix:temporal_tampering}
We consider three temporal manipulations: \textbf{Frame Drop}, \textbf{Frame Insert}, and \textbf{Frame Swap}. The first two change the number of frames, whereas the third changes only their order. For a video with \(f\) frames, the operations are defined as follows.

\begin{itemize}
\item \textbf{Frame Drop} removes a frame at a randomly selected temporal index \( p \), where \( p \in \{0, 1, \dots, f-1\} \). The tampered sequence is defined as:
\[
\begin{aligned}
\text{Drop}(\text{frame}_p) \to 
\{&\text{frame}_0, \dots, \text{frame}_{p-1}, \\
  &\text{frame}_{p+1}, \dots, \text{frame}_{f-1}\}.
\end{aligned}
\]

\item \textbf{Frame Insert} duplicates a neighboring frame and inserts it at a randomly selected index \( p \in \{1, 2, \dots, f-1\} \). Assuming the duplicated frame is the preceding frame, the tampered sequence becomes:
\[
\begin{aligned}
\text{Insert}(\text{frame}_{p-1}) \to 
\{&\text{frame}_0, \dots, \text{frame}_{p-1}, \\
& \text{frame}_{p-1}, \text{frame}_p, \dots, \text{frame}_{f-1}\}.
\end{aligned}
\]

\item \textbf{Frame Swap} exchanges the positions of temporally adjacent frames at regular intervals. Specifically, for
\[
p = 2 + 4k, \quad k \sim \left\{0, 1, 2, \dots, \left\lfloor \frac{f - 3}{4} \right\rfloor \right\},
\]
the swap operation is defined as:
\[
\text{Swap}(\text{frame}_p, \text{frame}_{p+1}) \rightarrow \{\dots, \text{frame}_{p+1}, \text{frame}_p, \dots\}.
\]

\end{itemize}

\subsection{Spatial Tampering Implementation}
\label{appendix:spatial_tampering}

We provide the details of spatial tampering in this subsection. We consider \textbf{Gaussian Blur}, \textbf{Colour Jitter}, and \textbf{Resolution Compression}. These operations are applied to every frame. Given a video frame \( \mathbf{F} \) with height \( H \) and width \( W \), their implementations are as follows:

\begin{itemize}
    \item \textbf{Gaussian Blur} applies a convolution with a Gaussian kernel \( K \) of size \( k \times k \), where \( k \) is a positive odd integer. The blurred frame \( \mathbf{F}' \) is computed as:
    \[
    \mathbf{F}' = \mathbf{F} * K,
    \]
    where \( * \) denotes the convolution operation. We use a kernel size of \(11 \times 11\) in our experiments.

    \item \textbf{Colour Jitter} perturbs the brightness and contrast of the frame. The transformed frame \( \mathbf{F}' \) is computed as:
    \[
    \mathbf{F}' = \alpha \mathbf{F} + \beta,
    \]
    where \( \alpha \sim \mathcal{U}(1 - s, 1 + s) \) controls contrast, \( \beta \sim \mathcal{U}(-s, s) \) controls brightness, and \( s \) is the jitter strength hyperparameter. We set \( s = 0.1 \) in all experiments.

    \item \textbf{Resolution Compression} simulates compression artifacts by downscaling and then upscaling the frame. Given a downscale factor \( r \in (0, 1) \), the frame \( \mathbf{F}' \) is obtained as:
    \[
    \mathbf{F}' = \mathrm{Upscale}(\mathrm{Downscale}(\mathbf{F}, r), (H, W)),
    \]
    where \(\mathrm{Downscale}(\cdot, r)\) resizes the frame to dimensions \( (rH, rW) \) and \(\mathrm{Upscale}(\cdot, (H, W))\) resizes it back to the original size using bilinear interpolation. We set a downscale factor of \( r = 0.5 \).
\end{itemize}

\subsection{Watermark Extraction Details}
\label{sec:extraction_details}

\paragraph{Decoding, detection, and recovery.}
\rev{These functions assess different outcomes: payload-bit recovery, evidence of watermark presence, and temporal correspondence. An edited video may retain enough evidence for detection without permitting complete message or temporal recovery.} The soft observations in Section~\ref{subsec:extraction} preserve confidence information for the BP-OSD decoder; the PRC interface and algorithms are given in Appendix~\ref{prc_watermark}.

\paragraph{Monte Carlo detection.}
Equation~\eqref{eq:mc_detection} compares the observed message distance with distances to random reference messages. \rev{With the add-one correction, the minimum attainable p-value is $1/(N+1)$. We use $N=1000$ for the original evaluation and $N=20{,}000$ for the supplementary detection experiments. The default threshold $\tau=0.005$ was selected in preliminary experiments; Appendix~\ref{sec:detection_reliability} reports sensitivity at $\tau\in\{0.001,0.005,0.01\}$ and empirical false-positive rates.}

\paragraph{Temporal matching.}
TMM corrects permutation-induced temporal misalignment through independent message matching and sorting. Surviving frame messages can still be localized after insertion or deletion, but information removed by trimming or destructive resampling cannot be reconstructed. The ablation in Appendix~\ref{sec:component_ablation} confirms that the primary gain is under block reordering.

\rev{For example, recovered indices $[3,4,1,2]$ specify both the correspondence after a block swap and the permutation needed to restore order. TMM combines message matching with temporal recovery. VideoMark represents temporal positions through registered messages; explicit frame-index embedding instead encodes positions directly. Both representations use a recovery rule to associate decoded information with temporal positions. Our ablation isolates the contribution of TMM to recovery within VideoMark.}

% AE decision X3V72vRmBf (2026-11-22, updated 2026-11-23).
% Author confirmed on 2026-11-23 that Monte Carlo references undergo the
% same TMM search. This clarification does not introduce new experiments.
\paragraph{\aerev{Temporal matching and video-level calibration.}}
\label{sec:tmm_calibration}
\aerev{TMM involves candidate selection followed by sorting. Sorting already matched frame-message pairs leaves a permutation-invariant aggregate distance unchanged. Selecting the closest message from the registered sequence, however, can reduce the distance even under the null hypothesis of no watermark. Video-level calibration therefore includes this selection step. In our Monte Carlo evaluation, the random references undergo the same TMM search and video-level aggregation as the observed messages. The candidate-sequence length, message length, and frame count are held fixed within each comparison.}

\aerev{Let $S_{\mathrm{obs}}$ denote the aggregate video-level distance after matching, and let $S_n$ denote the corresponding statistic from the $n$th random-reference trial, with smaller values indicating stronger agreement. The video-level counterpart of Equation~\eqref{eq:mc_detection} is
\begin{equation}
\label{eq:video_mc_detection}
 p_{\mathrm{video}}=
 \frac{1+\sum_{n=1}^{N}\mathbb{I}[S_n\leq S_{\mathrm{obs}}]}{N+1},
 \qquad d_{\mathrm{video}}=\mathbb{I}[p_{\mathrm{video}}<\tau].
\end{equation}
Both sides of the comparison thus use statistics computed after matching; a selected best match is not compared against an unmatched single-message reference. The entire video is the unit of the test, so the decision uses the aggregate statistic rather than treating frame-level tests as independent. The video-level FPR and false-acceptance controls in Appendix~\ref{sec:detection_reliability} provide empirical checks of this procedure under the evaluated settings.}

\subsection{Payload--Robustness Trade-off}
\label{sec:payload_robustness}
The robustness columns in Table~\ref{tab:main_result} use 32 bits per frame over 16 frames, matching VideoShield's total 512-bit payload. We additionally test the selected 512-bit-per-frame operating point in Table~\ref{tab:full_payload}. For the 16-frame ModelScope evaluation, these operating points correspond to 512 and 8,192 total bits, respectively. \rev{At 512 bits/frame, the H.264 accuracies are 0.5214, 0.5041, and 0.4998 for CRF18, CRF23, and CRF28, respectively.} The lower-payload H.264 results in Table~\ref{tab:realistic_temporal} are 0.6867, 0.6348, and 0.5220. The results rule out a payload-independent robustness claim: mild blur and brightness changes remain recoverable, whereas aggressive resizing and compression approach random bit accuracy. Table~\ref{tab:payload_curve} shows \rev{an overall decline under strong blur as payload grows, with some non-monotonic entries at intermediate blur strengths.}

\begin{table}[ht]
\centering
\caption{Spatial robustness on ModelScope at 512 bits per frame \rev{(8,192 total bits over 16 frames). This differs from the 32-bit-per-frame, 512-total-bit setting in Table~\ref{tab:realistic_temporal}.}}
\label{tab:full_payload}
\small
\begin{tabular}{lc@{\qquad}lc}
\toprule
Attack & Acc. & Attack & Acc. \\
\midrule
Identity & 1.0000 & Blur ($r=1$) & 0.9897 \\
Blur ($r=2$) & 0.9496 & Blur ($r=4$) & 0.8927 \\
Blur ($r=8$) & 0.6006 & Brightness ($0.5\times$) & 0.9899 \\
Brightness ($0.75\times$) & 0.9900 & Brightness ($1.25\times$) & 0.9899 \\
Brightness ($1.5\times$) & 0.9900 & Resize ($0.9\times$) & 0.4991 \\
Resize ($0.8\times$) & 0.4989 & Resize ($0.7\times$) & 0.5003 \\
Resize ($0.5\times$) & 0.4997 & H.264 (CRF18) & \rev{0.5214} \\
H.264 (CRF23) & \rev{0.5041} & H.264 (CRF28) & \rev{0.4998} \\
\bottomrule
\end{tabular}
\end{table}

\begin{table}[ht]
\centering
\caption{Gaussian-blur accuracy versus per-frame payload.}
\label{tab:payload_curve}
\small
\begin{tabular}{ccllll}
\toprule
Bits & Identity & $r=1$ & $r=2$ & $r=4$ & $r=8$ \\
\midrule
32  & 1.0000 & 1.0000 & 1.0000 & 1.0000 & 0.9741 \\
64  & 1.0000 & 1.0000 & 0.9995 & 0.9663 & 0.8979 \\
128 & 1.0000 & 1.0000 & 0.9917 & 0.9415 & 0.8501 \\
256 & 1.0000 & 1.0000 & 0.9333 & 0.9040 & 0.7419 \\
512 & 1.0000 & 0.9897 & 0.9496 & 0.8927 & 0.6006 \\
\bottomrule
\end{tabular}
\end{table}

\subsection{Extraction Assumptions and Latent Attacks}
\label{sec:extraction_assumptions}
VideoMark is non-blind with respect to its model and watermark resources. The verifier uses the corresponding UNet and VAE weights, scheduler, inversion steps, guidance scale, and preprocessing settings, together with the private PRC key and registered master sequence. \rev{The original generation prompt and conditioning image are not required: T2V uses an empty prompt, and I2V uses the fixed literal string \texttt{\char34 none\char34} and the first frame of the video being verified. These conditions also apply to the reported mismatch experiments.} Moderate step and guidance mismatches leave accuracy unchanged in our test (Table~\ref{tab:mismatch_latent}), because decoding primarily depends on the recovered sign confidence. This tolerance does not extend to all pipeline changes: VAE-space smoothing or denoising destroys the sign structure, while latent quantization is less harmful.

\begin{table}[ht]
\centering
\caption{Mismatched extraction and VAE-space post-processing.}
\label{tab:mismatch_latent}
\small
\begin{tabular}{llc}
\toprule
Category & Setting or attack & Acc. \\
\midrule
Matched & 25 steps, guidance 9 & 1.0000 \\
Step mismatch & 20 / 30 inversion steps & 1.0000 / 1.0000 \\
Guidance mismatch & 7.5 / 12 & 1.0000 / 1.0000 \\
\midrule
Latent attack & Gaussian blur ($k=3$ / $k=5$) & 0.4998 / 0.5004 \\
Latent attack & Denoising & 0.5005 \\
Latent attack & Quantization (4-bit / 2-bit) & 0.9728 / 0.9666 \\
\bottomrule
\end{tabular}
\end{table}

\subsection{Long-Video Evaluation}
Table~\ref{tab:long_video} compares both inversion-based methods as length increases. Both degrade gradually, so VideoMark's length-independent registration should not be interpreted as length-independent extraction accuracy. Segment-wise watermarking is a possible mitigation for longer videos, \rev{but is not evaluated here.}
\begin{table}[ht]
\centering
\caption{Bit accuracy versus video length.}
\label{tab:long_video}
\small
\begin{tabular}{cccccc}
\toprule
Frames & 16 & 24 & 32 & 48 & 64 \\
\midrule
VideoMark & 1.0000 & 0.9941 & 0.9504 & 0.9249 & 0.9091 \\
VideoShield & 1.0000 & 0.9939 & 0.9602 & 0.9458 & 0.9315 \\
\bottomrule
\end{tabular}
\end{table}

\subsection{Component Ablation}
\label{sec:component_ablation}
Removing PRC reduces clean and attacked accuracy to chance. TMM provides its largest gain under block swapping, from 0.6473 to 1.0000, but yields little improvement when temporal information is irreversibly removed or resampled. Without frame-wise encoding, changed-length inputs cannot be decoded because the latent sequence dimension no longer matches the registered watermark.

\rev{The ablation isolates each component's contribution to message recovery within VideoMark; Appendix~\ref{sec:detection_reliability} reports the complementary video-level detection results.}

\begin{table}[ht]
\centering
\caption{Component ablations. A dash denotes an undecodable sequence-length mismatch.}
\label{tab:component_ablation}
\small
\begin{tabular}{lcccc}
\toprule
Setting & Identity & Block swap & Speed $2\times$ & Trim \\
\midrule
Full model & 1.0000 & 1.0000 & 0.5950 & 0.5015 \\
w/o PRC & 0.4991 & 0.5014 & 0.5014 & 0.5011 \\
w/o TMM & 1.0000 & 0.6473 & 0.5540 & 0.5024 \\
w/o frame-wise & 1.0000 & 1.0000 & -- & -- \\
\bottomrule
\end{tabular}
\end{table}

\subsection{Runtime, Memory, and Matching Scalability}
\label{sec:runtime}
Across 20 16-frame videos with 25 DDIM steps, embedding and extraction take $11.25\pm4.56$ and $26.40\pm1.09$ seconds (12.30/10.17GB peak); BP-OSD costs $1.626\pm0.060$ seconds per frame. Expanding one master sequence from 300 to 30,000 entries changes extraction from $26.90\pm0.22$ to $28.03\pm0.21$ seconds. This tests sequence matching, not multi-key video-database search, which remains unevaluated; non-blind extraction remains costly.

% Source: October 21 author responses S7h7VmhViW, lpiTiPemTd,
% kBHMkOb5AE, and 6xdRw4p7B9. Values are reported experiments, not reruns.
% Author's latest clarification (2026-11-02): each attack was evaluated
% on 100 watermarked positives and 100 unwatermarked negatives, with
% 0/100 false positives. This supersedes the earlier negative-sample reply.
\subsection{\rev{Detection Reliability and Threshold Sensitivity}}
\label{sec:detection_reliability}
\begin{revision}
The supplementary ModelScope detection experiments use 32 bits per frame (512 bits in the original 16-frame video) and $N=20{,}000$ Monte Carlo reference samples. We retain the empirically selected operating threshold $\tau=0.005$ and examine 0.001 and 0.01 to assess sensitivity. We report video-level acceptance rates alongside the bit-accuracy and temporal-index recovery evaluations.

Table~\ref{tab:detection_controls} evaluates 100 unattacked videos per setting, with the same videos reused across thresholds within each setting. The 95\% Wilson intervals use videos as the statistical units.
\end{revision}

\begin{table}[H]
\centering
\revisioncolor
\caption{\rev{Unattacked detection controls on ModelScope. Entries are accepted/tested videos. Each row has identical counts at all three thresholds, so one 95\% Wilson interval is shown. Wrong-key and wrong-sequence rows measure false acceptance.}}
\label{tab:detection_controls}
\small
\begin{tabular}{p{0.36\linewidth}lccc l}
\toprule
Setting & Metric & $\tau=.001$ & $.005$ & $.01$ & 95\% CI \\
\midrule
Unwatermarked videos & FPR & 0/100 & 0/100 & 0/100 & [0\%, 3.70\%] \\
Watermarked, incorrect key & FAR & 0/100 & 0/100 & 0/100 & [0\%, 3.70\%] \\
Watermarked, incorrect sequence & FAR & 0/100 & 0/100 & 0/100 & [0\%, 3.70\%] \\
Watermarked, correct key and sequence & TPR & 100/100 & 100/100 & 100/100 & [96.30\%, 100\%] \\
\bottomrule
\end{tabular}
\end{table}

\begin{revision}
Table~\ref{tab:detection_thresholds} evaluates 100 watermarked videos and 100 unwatermarked videos under each attack condition and reports TPR confidence intervals at $\tau=0.005$. FPR is the fraction of unwatermarked videos incorrectly accepted as watermarked. At all three thresholds, each attack yields 0/100 false positives (observed FPR 0\%; 95\% Wilson interval [0\%, 3.70\%]). TPR changes by at most four percentage points within an attack condition, indicating stable detection performance around the selected threshold.
\end{revision}

\begin{table}[H]
\centering
\revisioncolor
\caption{\rev{VideoMark detection under editing on ModelScope, using 100 watermarked and 100 unwatermarked videos per attack at 32 bits/frame. TPR is in percent; the count and 95\% Wilson interval correspond to $\tau=0.005$. FPR is 0/100 at every listed threshold and attack (95\% CI [0\%, 3.70\%]).}}
\label{tab:detection_thresholds}
\small
\begin{tabular}{lccccl}
\toprule
Attack & $\tau=.001$ & $.005$ & $.01$ & Detected/tested & 95\% CI \\
\midrule
Identity & 100 & 100 & 100 & 100/100 & [96.30\%, 100\%] \\
Head trim (25\%) & 100 & 100 & 100 & 100/100 & [96.30\%, 100\%] \\
Middle trim (25\%) & 100 & 100 & 100 & 100/100 & [96.30\%, 100\%] \\
Tail trim (25\%) & 100 & 100 & 100 & 100/100 & [96.30\%, 100\%] \\
Speed ($0.5\times$) & 100 & 100 & 100 & 100/100 & [96.30\%, 100\%] \\
Speed ($2\times$) & 100 & 100 & 100 & 100/100 & [96.30\%, 100\%] \\
FPS (10$\rightarrow$24) & 100 & 100 & 100 & 100/100 & [96.30\%, 100\%] \\
FPS (10$\rightarrow$30) & 100 & 100 & 100 & 100/100 & [96.30\%, 100\%] \\
Block swap & 91 & 92 & 92 & 92/100 & [85.00\%, 95.89\%] \\
Frame averaging & 52 & 54 & 56 & 54/100 & [44.26\%, 63.44\%] \\
Cross-fade & 100 & 100 & 100 & 100/100 & [96.30\%, 100\%] \\
H.264 (CRF18) & 100 & 100 & 100 & 100/100 & [96.30\%, 100\%] \\
H.264 (CRF23) & 99 & 99 & 99 & 99/100 & [94.55\%, 99.82\%] \\
H.264 (CRF28) & 64 & 65 & 67 & 65/100 & [55.25\%, 73.64\%] \\
\bottomrule
\end{tabular}
\end{table}

\begin{revision}
Table~\ref{tab:detection_vs_videoshield} compares video-level detection with VideoShield using 100 watermarked and 100 unwatermarked ModelScope videos per attack for each method. VideoMark uses the Monte Carlo threshold $\tau=0.01$, and VideoShield uses a bit-accuracy threshold of 0.60. Every reported 0\% FPR corresponds to 0/100 false positives, with a 95\% Wilson interval of [0\%, 3.70\%]. Dashes denote changed-length inputs unsupported by the evaluated VideoShield decoder.

VideoMark has higher observed TPR under frame averaging and cross-fading, while VideoShield has higher TPR under block swapping and stronger H.264 compression. After trimming, surviving watermark evidence supports detection even when complete message recovery is impaired. Appendix~\ref{sec:component_ablation} separately evaluates the contribution of TMM to temporal recovery.
\end{revision}

\begin{table}[H]
\centering
\revisioncolor
\caption{\rev{Detection comparison on ModelScope (rates in percent), using 100 watermarked and 100 unwatermarked videos per attack for each method. The original payload is 512 bits for both methods; VideoMark uses 32 bits/frame. VideoMark uses $\tau=0.01$, whereas VideoShield uses bit-accuracy threshold 0.60. Each reported FPR of 0 corresponds to 0/100 false positives. Frame counts after editing are given in Table~\ref{tab:realistic_temporal}.}}
\label{tab:detection_vs_videoshield}
\small
\begin{tabular}{lcccc}
\toprule
& \multicolumn{2}{c}{VideoMark} & \multicolumn{2}{c}{VideoShield} \\
\cmidrule(lr){2-3}\cmidrule(lr){4-5}
Attack & TPR & FPR & TPR & FPR \\
\midrule
Identity & 100 & 0 & 100 & 0 \\
Head trim (25\%) & 100 & 0 & -- & -- \\
Middle trim (25\%) & 100 & 0 & -- & -- \\
Tail trim (25\%) & 100 & 0 & -- & -- \\
Speed ($0.5\times$) & 100 & 0 & -- & -- \\
Speed ($2\times$) & 100 & 0 & -- & -- \\
FPS (10$\rightarrow$24) & 100 & 0 & -- & -- \\
FPS (10$\rightarrow$30) & 100 & 0 & -- & -- \\
Block swap & 92 & 0 & 100 & 0 \\
Frame averaging & 56 & 0 & 0 & 0 \\
Cross-fade & 100 & 0 & 15 & 0 \\
H.264 (CRF18) & 100 & 0 & 100 & 0 \\
H.264 (CRF23) & 99 & 0 & 100 & 0 \\
H.264 (CRF28) & 67 & 0 & 100 & 0 \\
\bottomrule
\end{tabular}
\end{table}

\FloatBarrier
\subsection{Parameter Sensitivity}
\label{sec:parameter_sensitivity}
This subsection supplements the main-text video-length and payload analyses with extended length, spatial-resolution, and inversion-step evaluations.

\subsubsection{Extended Video Length Evaluation}

The protocol accepts different frame counts without length-specific registration, but extraction accuracy is not length invariant. In the expanded T2V evaluation, accuracy decreases from 1.0000 at 16 frames to 0.9941, 0.9504, 0.9249, and 0.9091 at 24, 32, 48, and 64 frames, respectively (Appendix Table~\ref{tab:long_video}). VideoShield shows a similar trend, indicating a shared limitation of inversion-based methods rather than evidence of arbitrary-length robustness. Segment-wise watermarking may limit accumulated inversion error, but we do not evaluate that mitigation here.

\subsubsection{Impact of Spatial Resolution} Table~\ref{tab:frame_resolution} shows that resolution sensitivity is model dependent. ModelScope remains between 0.998 and 1.000 across all tested resolutions, whereas I2VGen-XL peaks at 0.997 at $512\times512$ and decreases to 0.980 and 0.973 at $960\times544$ and $1280\times720$. The absence of a monotonic decline for ModelScope cautions against attributing the effect to resolution alone; the interaction among VAE reconstruction, resizing, and inversion error differs across backbones. Accordingly, the $512\times512$ operating point should be understood as the selected configuration for these experiments, not as a universally optimal watermark resolution.
\begin{table}[H]
\centering
\caption{Evaluation of bit accuracy across varying spatial resolutions.}
%\begin{adjustbox}{width=\columnwidth,center}
\begin{tabular}{lcccc}
\toprule
\multirow{2}{*}{Model} & \multicolumn{4}{c}{Spatial Resolution} \\ \cmidrule(lr){2-5}
 & \(256\times256\) & \(512\times512\) & \(960\times544\) & \(1280\times720\) \\
\midrule
Modelscope & 1.000 & \cellcolor{myblue}\textbf{1.000} & 1.000 & 0.998 \\
I2VGen-XL  & 0.996 & \cellcolor{myblue}\textbf{0.997} & 0.980 & 0.973 \\
\bottomrule
\end{tabular}
%\end{adjustbox}
\label{tab:frame_resolution}
\end{table}

\subsubsection{Impact of Inversion Steps} Table~\ref{tab:ablation_steps} reveals a marked backbone asymmetry. ModelScope remains at or above 0.986 across the tested inference/inversion grid and often reaches 1.000 even when the step counts differ. I2VGen-XL is more configuration sensitive: its best reported value, 0.997, occurs at the matched 50/50 setting, while several mismatched settings fall to 0.892--0.957. More steps are not uniformly better, so the matrix reflects scheduler--backbone interactions rather than a simple accuracy--compute curve. We use 25 steps for T2V and 50 for I2V in the main experiments because these settings combine high matched accuracy with their respective generation configurations; this choice does not remove the non-blind requirement to know or approximate the extraction pipeline.
\begin{table}[H]
\centering
\caption{Bit accuracy across different inference (Inf.) and inversion (Inv.) steps. The main experimental configuration is marked.}
%\begin{adjustbox}{width=\columnwidth, center}
\begin{tabular}{c *{4}{c} *{4}{c}}
\toprule
\multirow{2}{*}{\diagbox[width=4em]{\textbf{Inf.}}{\textbf{Inv.}}}
    & \multicolumn{4}{c}{\textbf{ModelScope}} 
    & \multicolumn{4}{c}{\textbf{I2VGen-XL}} \\ 
\cmidrule(lr){2-5} \cmidrule(lr){6-9}
& 10 & 25 & 50 & 100 & 10 & 25 & 50 & 100 \\
\midrule
10  & 0.986 & 0.995 & 0.991 &  1.000 
    & 0.922 & 0.903 & 0.892 &  0.915 \\
25  & 0.998 &  \cellcolor{myblue}\textbf{1.000} &  1.000 &  1.000 
    & 0.940 &  0.985 &  0.982 &  0.989 \\
50  & 0.999 &  1.000 & 0.999 &  1.000 
    & 0.957 & 0.981  & \cellcolor{myblue}\textbf{0.997} &  0.993 \\
100 &  1.000 &  1.000 &  1.000 &  1.000 
    &  0.957 &  0.987 &  0.988 &  0.993 \\
\bottomrule
\end{tabular}
%\end{adjustbox}

\label{tab:ablation_steps}
\end{table}

\FloatBarrier

\end{document}